\documentclass[12pt]{iopart}

\usepackage{iopams}
\usepackage{graphicx}

\begin{document}

\title[Magneto-optical conductivity in Graphene]{Magneto-optical conductivity in Graphene}

\author{V P Gusynin$^1$, S G Sharapov$^2$ and J P Carbotte$^2$}

\address{$^1$ Bogolyubov Institute for Theoretical Physics, 14-b
        Metrologicheskaya Street, Kiev, 03143, Ukraine}
\address{$^2$ Department of Physics and Astronomy, McMaster University,
        Hamilton, Ontario, Canada, L8S 4M1}
        \ead{vgusynin@bitp.kiev.ua} \ead{carbotte@mcmaster.ca}  \ead{sharapov@bitp.kiev.ua}

\begin{abstract}
Landau level quantization in graphene reflects the Dirac nature of
its quasiparticles and has been found to exhibit an unusual integer
quantum Hall effect. In particular the lowest Landau level can be
thought as shared equally by electrons and holes and this leads to
characteristic behaviour of the magneto-optical conductivity as a
function of frequency $\Omega$ for various values of the chemical
potential $\mu$. Particular attention is paid to the optical
spectral weight under various absorption peaks and its
redistribution as $\mu$ is varied. We also provide results for
magnetic field $B$ as well as chemical potential sweeps at selected
fixed frequencies which can be particularly useful for  possible
measurements in graphene. Both diagonal and Hall conductivity are
considered.
\end{abstract}

\submitto{\JPCM}
\maketitle

\section{Introduction}
\label{sec:intro}

Recent experimental studies \cite{Geim2005Nature,Kim2005Nature} of
the dynamics of electrons and holes in graphene (a single atomic
layer of graphite \cite{Novoselov2004Science}) have revealed unusual
behaviour related to the Dirac nature of its quasiparticles. Two
dimensional graphene has a honeycomb lattice structure with two
atoms per unit cell. Its band structure consists of two inequivalent
pairs of cones with apex at the Brillouin zone corners. For zero
chemical potential the lower energy cones are completely filled and
the upper empty. In a graphene device an applied gate voltage can be
used to introduce electrons in the upper band or, by voltage
reversal, holes in the lower band (cones).

The quasiparticles in graphene obey Dirac
\cite{Semenoff1984PRL,DiVincenzo1984PRB} rather than the
Schr\"odinger equation and this has profound implications for their
dynamics. The unconvetional quantum Hall effect was expected
theoretically
\cite{Zheng2002PRB,Gusynin2005PRL,Gusynin2006PRB,Peres2006PRB,Neto2006PRB}
and recently observed \cite{Geim2005Nature,Kim2005Nature} to have
half (divided by spin and valley degeneracy) rather than integer
filling factors. The predicted phase shift of $\pi$ in the de Haas
van Alphen \cite{Sharapov2004PRB,Luk'yanchuk2004PRL,Mikitik1999PRL}
and Shubnikov de Haas \cite{Gusynin2005PRB} oscillations was also
seen experimentally \cite{Geim2005Nature,Kim2005Nature}. Another
feature related to the Dirac-like character of the carriers in
graphene seen in the dc measurements is a finite effective cyclotron
mass for the massless Dirac quasiparticles which varies as the
square root of the number of carriers
\cite{Sharapov2004PRB,Gusynin2005PRB,Geim2005Nature,Kim2005Nature}.

In this paper we consider the magneto-optical conductivity of
graphene. Work without a magnetic field includes the calculations of
Ando, Zheng and Suzuura \cite{Ando2002JPSJ} who considered the
effect on the frequency dependent conductivity of short- and
long-range scatterers in a self consistent Born approximation. More
recent work  \cite{Gusynin2006micro} describes several anomalous
properties of the microwave conductivity of graphene with, as well
as without magnetic field. These properties are directly related to
the Dirac nature of the quasiparticles. Several analytic formulas
for the longitudinal as well as Hall ac conductivity are given in
the paper  \cite{Gusynin2006PRB}. They also present extensive
results for dc properties and preliminary data on the real part of
$\sigma_{xx}(\Omega)$ vs $\Omega$ in the optical region. Another
extensive work by Peres et al. \cite{Peres2006PRB}  on the ac
conductivity in graphene treats localized impurities in a
self-consistent fashion  as well as extended edge and grain
boundaries including also effects of electron-electron interactions
and self-doping.

In this paper we follow most closely Ref.~\cite{Gusynin2006PRB}
which we extend in several directions. An aim is to provide simpler
analytic formulae which should prove useful in the analysis of
experiment and check on their accuracy. Another is to consider
magnetic field as well as chemical potential sweeps possible in
graphene field effect transistor devices
\cite{Geim2005Nature,Kim2005Nature,Novoselov2004Science,Bunch2005Nano}.
Recent work by Li et al. in organic metals \cite{Basov-organic} has
demonstrated that the ac measurements are also possible in such
devices. Consistent with our aim, we consider impurities only in
simplified scattering rate model and neglect real part
renormalizations although these could easily be included if  wished.
While the renormalization effects beyond those included in this
simplified model might become important for the interpretation of
future experiment, we point out that so far the free quasiparticle
model with associated transport lifetime has been remarkably
successful in understanding the dc results of
Refs.~\cite{Geim2005Nature,Kim2005Nature}.

In Sec.~\ref{sec:cond-general} we relate the magneto-optical
conductivity tensor to the Dirac fermionic Green's function through
Kubo formula. The general formulas obtained can be greatly
simplified and closed form expressions are obtained in two cases. In
general the fermionic self-energy can depend on energy and Landau
level index as well as temperature and value of the external
magnetic field $B$. Under the assumption that variations with Landau
level index $n$ can be neglected, the sum over transitions between
neighboring Landau levels can be carried out explicitly and a closed
form expression is obtained for conductivity in terms of digamma
function. A single integral over an internal frequency remains. The
expression obtained is suitable for calculations of the ac
conductivity for any value of temperature, chemical potential and
magnetic field. Its microwave frequency limit has been used to
describe properties of graphene in Ref.~\cite{Gusynin2006micro}. In
the case when energy dependence of the fermionic self-energy can be
neglected, the internal integration over energy can be done and what
remains is the sum over the Landau level index $n$ of Lorentzian
forms multiplied by thermal factors and algebraic weighting factors.
The weighting factors depend on the Landau level energies as well as
the excitonic gap (see e.g.
Refs.~\cite{Khveshchenko2001PRL,Khveshchenko2001aPRL,Gorbar2002PRB})
should one wish to include this possibility. When we compare
numerical results obtained from Lorentzian model and from the
previous more general expressions in the limit of constant
scattering rate, we find good quantitative agreement between the
two. This provides support for the analysis of experimental data
\cite{Li2006}. In Sec.~\ref{sec:low-B} we consider the low field
limit for the Lorentzian model derived in Sec.~\ref{sec:Lorentzian}
and establish its correspondence with previously known results. In
Sec.~\ref{sec:cond-diagonal} we present the numerical results for
the real part of the diagonal conductivity as a function of photon
energy $\Omega$ for fixed value of magnetic field and various values
of  chemical potential. We also provide results for fixed photon
energy while sweeping either chemical potential or external magnetic
field which should prove useful in comparing with experiments. The
effect of opening of an excitonic gap on the absorption lines is
described. In Sec.~\ref{sec:weight} a discussion of optical spectral
weight redistribution by the magnetic field is given. Section
\ref{sec:cond-Hall} is structured in parallel to
Sec.~\ref{sec:cond-diagonal} but deals with the absorptive part of
the off-diagonal Hall magneto-optical conductivity. Discussion and
conclusions are given in Sec.~\ref{sec:concl}. Some of the algebra
needed in this work is found in an \ref{sec:A}.

\section{Analytic expressions for optical conductivity}
\label{sec:cond-general}

The optical conductivity tensor is calculated using the Kubo formula
\begin{equation}
\label{Kubo-cond} \sigma_{ij} (\Omega)=
\frac{\Pi^R_{ij}(\Omega+i0)}{i \Omega},
\end{equation}
where $\Pi^R_{ij}(\Omega)$ is the retarded current-current
correlation function which in the bubble approximation is given by
\begin{equation}
\label{electric_cond} \fl
\Pi_{ij}(\Omega+i0)={e^2v_F^2}\int_{-\infty}^\infty d\omega
d\omega^\prime\frac{n_F(\omega^\prime)-n_F(\omega)}
{\omega-\omega^\prime-\Omega-i0}\int\frac{d^2k}{(2\pi)^2}{\rm
tr}\left[\gamma^iA(\omega,\mathbf{k})\gamma^j
A(\omega^\prime,\mathbf{k})\right], \quad i =1,2
\end{equation}
where $n_F(\omega)$ is the Fermi distribution function $n_F(\omega)=
1/[\exp((\omega-\mu)/T)+1]$, $\mbox{tr}$ not only takes care of the
$4\times 4$ $\gamma$ matrices [$\gamma^\nu = \sigma_3\otimes
(\sigma_3, i \sigma_2, -i\sigma_1)$], but includes also the
summation over flavour (spin) index. Here
\begin{equation}
\label{spect-fun-magfield} \fl A(\omega
,{\mathbf{k}})=e^{-\frac{c{\mathbf{k}}^{2}}{|eB|}}
\sum_{n=0}^{\infty }\frac{(-1)^{n} \Gamma_n(\omega)}{2 \pi
M_{n}}\left[ \frac{(\gamma
^{0}M_{n}+\Delta)f_{1}({\mathbf{k}})+f_{2}({\mathbf{k}})}{(\omega
-M_{n})^{2}+\Gamma_n^{2}(\omega)}+\frac{(\gamma ^{0}M_{n}-\Delta
)f_{1}({\mathbf{k}} )-f_{2}({\mathbf{k}})}{(\omega
+M_{n})^{2}+\Gamma_n^{2}(\omega)}\right]
\end{equation}
is the spectral function (decomposed over Landau levels) associated
with the translation invariant part of the Dirac fermion Green's
function in an external magnetic field $\mathbf{B}$ applied
perpendicular to the plane along the positive $z$ direction (see
e.g. Refs.~\cite{Gorbar2002PRB,Sharapov2004PRB}). In
Eq.~(\ref{spect-fun-magfield})
\begin{equation}
\label{expr:f1-f2} \fl f_{1}({\mathbf{k}})=2\left[ P_{-}L_{n}\left(
\frac{2c{\mathbf{k}}^{2}}{|eB|}\right) -P_{+}L_{n-1}\left(
\frac{2c{\mathbf{k}}^{2}}{|eB|}\right) \right] ,\quad f_{2}({
\mathbf{k}})=4v_F{\mathbf{k}}{\pmb{\gamma}}L_{n-1}^{1}\left(
\frac{2c{\mathbf{k}}^{2}}{|eB|} \right)
\end{equation}
with $P_{\pm }=(1\pm i\gamma ^{1}\gamma ^{2}{\rm sgn}(eB))/2$ being
projectors and $L^{\alpha}_{n}(z)$ the generalized Laguerre
polynomials. By definition, $L_{n}(z)\equiv L^{0}_{n}(z)$ and
$L^{\alpha}_{-1}(z)\equiv 0$. The energies of the relativistic
Landau levels in Eq.~(\ref{spect-fun-magfield}) are
\begin{equation}
\label{M_n} E_n=\pm M_n, \qquad
M_{n}=\sqrt{\Delta^{2}+2nv_F^2|eB|/c},
\end{equation}
where the energy scale associated with the magnetic field expressed
in the units of temperature reads
\begin{equation}
\frac{e B v_F^2}{c} \to \frac{eB \hbar v_F^2}{c} \frac{1}{k_B^2} =
8.85 \times 10^{-8} \mbox{K}^2 v_F^2(\mbox{m/s}) B(T),
\end{equation}
where $v_F$ is the Fermi velocity in graphene given in m/s and the
field $B$ is given in Tesla. In the following we set $\hbar =
k_B=1$, and in some places $e=c=1$, unless stated explicitly
otherwise. For the numerical calculations we use the value $v_F
\approx 10^6 \mbox{m/s}$ \cite{Geim2005Nature,Kim2005Nature}  which
leads to the relationship $eB \to (8.85 \times 10^4 \mbox{K}^2)
B(\mbox{T})$. We consider relatively low fields $B \lesssim 17
\mbox{T}$, where spin splitting is unresolved  \cite{Zhang2006PRL},
so that we can assume that the above mentioned summation over flavor
index simply gives $N_f=2$ in all expressions below. The Landau
level energies (\ref{M_n}) include also an excitonic gap $\Delta$.
The physical meaning of this singlet excitonic gap  is directly
related to the electron density imbalance between the A and B
sublattices of the bi-particle hexagonal lattice of graphene
\cite{Khveshchenko2001PRL,Khveshchenko2001aPRL} and there are strong
indications \cite{Gusynin2006catalysis} that it was indeed observed
in recent experiments \cite{Zhang2006PRL,Geim-private}. We will see
here that optical measurements done on graphene can be very useful
in investigations of the excitonic gap.

Finally, the scattering rate $\Gamma_n(\omega)$ is expressed via the
retarded fermion self-energy, $\Gamma_{n}(\omega)= -{\rm
Im}\Sigma_n^R(\omega)$ which in general depends on the energy,
temperature, field and the Landau levels index $n$. This
self-energy, which in general has also a real part, has to be
determined self-consistently from the Schwinger-Dyson equation. This
equation can be solved analytically \cite{Khveshchenko2001aPRL} and
numerically as was done in Refs.~\cite{Zheng2002PRB,Peres2006PRB}.
In our paper we consider $\Gamma_{n}(\omega)$ as a phenomenological
parameter for the two cases: (i) $\Gamma(\omega)=\Gamma_n(\omega)$
is independent of the Landau level index $n$ and (ii)
$\Gamma_n=\Gamma_n(\omega)$ is independent of the energy $\omega$.
Under these assumptions the optical conductivity can be studied
analytically.

\subsection{Frequency dependent scattering rate}

Assuming that $\Gamma_n(\omega)$ is independent of the Landau level
index, i.e. $\Gamma(\omega)= \Gamma_n(\omega)$, one can calculate
the sum over Landau levels and express the diagonal ac conductivity
in the closed form \cite{Gusynin2006PRB}
\begin{eqnarray}
\label{optical-diagonal}
\fl \mbox{Re} \, \sigma_{xx}(\Omega)
=\frac{e^2N_f}{4\pi^2\,\Omega}\int_{-\infty}^\infty d\omega
[n_F(\omega)-n_F(\omega^\prime)]{\rm Re}\left\{\frac{2B}{\Delta^2-
(\tilde\omega+i\Gamma)^2}\left[\Xi_{1}(-B)-\Xi_{2}(-B) \right] \right.\\
\fl+ \left. \left[\Xi_{1}(-B)+\Xi_{1}(+B)-\Xi_{2}(-B)-\Xi_{2}(+B)
\right]\psi\left(\frac{\Delta^2-(\tilde
\omega+i\Gamma)^2}{2B}\right)+(\tilde \omega\leftrightarrow \tilde
\omega^\prime, \Gamma\leftrightarrow\Gamma^\prime) \right\}
\nonumber.
\end{eqnarray}
Here $\psi$ is the digamma function, we denoted $B\equiv
v_F^2|eB|/c$, and also included the renormalization of energy caused
by the real part of self-energy $\tilde \omega(\omega) = \omega -
\mbox{Re} \, \Sigma(\tilde \omega)$, $\tilde \omega^\prime=\tilde
\omega (\omega+\Omega)$, $\Gamma=\Gamma(\tilde
\omega),\Gamma^\prime=\Gamma(\tilde \omega^\prime)$ and introduced
the following short-hand notations
\begin{eqnarray}\label{Xi}
\fl \Xi_{1}(\pm B) \equiv \Xi_{1}(\tilde \omega,\tilde
\omega^\prime,\Gamma,\Gamma^\prime,\pm B)=
\frac{(\tilde\omega^\prime+i\Gamma^\prime)(\tilde\omega+i\Gamma)-\Delta^2}
{[\tilde\omega-\tilde\omega^\prime+i(\Gamma-\Gamma^\prime)][\tilde\omega+\tilde\omega^\prime+i(\Gamma+\Gamma^\prime)]\pm 2B},
\nonumber \\
\fl \Xi_{2}(\pm B) \equiv
\Xi_{2}(\tilde\omega,\tilde\omega^\prime,\Gamma,\Gamma^\prime,\pm
B)=
\frac{(\tilde\omega^\prime-i\Gamma^\prime)(\tilde\omega+i\Gamma)-\Delta^2}
{[\tilde\omega-\tilde\omega^\prime+i(\Gamma+\Gamma^\prime)][\tilde\omega+\tilde\omega^\prime+
i(\Gamma-\Gamma^\prime)] \pm 2B}.
\end{eqnarray}
The advantage of Eq.~(\ref{optical-diagonal}) is that the $\psi$
function contains the contribution to conductivity from {\em all}
transitions between neighboring Landau levels. An infinite number of
these transitions need to be taken into account when the limit
$B\to0$ is considered, so that the zero field limit is easily
treatable \cite{Gusynin2006PRB} on the base of
Eq.~(\ref{optical-diagonal}). Another important feature of
Eq.~(\ref{optical-diagonal}) is that we kept the frequency dependent
impurity scattering rate $\Gamma(\omega)$ which allows us to
investigate its influence on the shape of the Drude peak
\cite{Gusynin2006micro}.

Although in our work we will mostly use Eq.~(\ref{optical-diagonal})
for the numerical computations of the diagonal conductivity, there
is a possibility to derive simple approximate expressions for the
diagonal and Hall conductivities which turn out to be very useful
when one is interested in the resonance peaks of these
conductivities in the infrared region.

\subsection{Landau level index dependent scattering rate and magneto-optical Lorentzian model}
\label{sec:Lorentzian}

For analyzing experimental data it is very useful to have a
magneto-optical Lorentzian model for the complex conductivity,
$\sigma_{\pm}(\Omega) = \sigma_{xx}(\Omega) \pm i
\sigma_{xy}(\Omega)$ due to inter- and intraband Landau level
transitions \cite{Lax1967}. For the case of Dirac fermions it is
considered in \ref{sec:A}, whereof we obtain the complex diagonal
conductivity
\begin{eqnarray}
\label{sigma_xx-complex-corrected}
\fl \sigma_{xx}(\Omega)=-\frac{e^2v_F^2|eB|}{2\pi ci}\\
\fl \sum_{n=0}^\infty
\left\{\left(1-\frac{\Delta^2}{M_nM_{n+1}}\right)\left([n_{F}(M_n) -
n_F(M_{n+1})] + [n_F(-M_{n+1}) - n_F(-M_{n})]\right)
\right. \nonumber \\
\fl\left.\times\left(\frac{1}{M_n - M_{n+1}+\Omega +i (\Gamma_n +
\Gamma_{n+1})}-\frac{1}{M_n - M_{n+1}-\Omega -i (\Gamma_n +
\Gamma_{n+1})}\right)\frac{1}{M_{n+1}-M_n}\right.\nonumber \\
\fl
+\left.\left(1+\frac{\Delta^2}{M_nM_{n+1}}\right)\left([n_{F}(-M_n)
- n_F(M_{n+1})] + [n_F(-M_{n+1}) -
n_F(M_{n})]\right)\frac{1}{M_{n+1}+M_n}\right. \nonumber \\
\fl \times\left.\left(\frac{1}{M_n +M_{n+1} +\Omega+ i(\Gamma_n +
\Gamma_{n+1})} -\frac{1}{M_n +M_{n+1} -\Omega- i(\Gamma_n +
\Gamma_{n+1})}\right)\right\}, \nonumber
\end{eqnarray}
and the complex Hall conductivity
\begin{eqnarray}
\label{sigma_xy-complex-corrected}
\fl \sigma_{xy}(\Omega)=\frac{e^2v_F^2eB}{2\pi c}
\sum_{n=0}^\infty\left([n_{F}(M_n) - n_F(M_{n+1})] -
[n_F(-M_{n+1}) - n_F(-M_{n})]\right)\nonumber \\
\fl \times
\left\{\left(1-\frac{\Delta^2}{M_nM_{n+1}}\right)\frac{1}{M_{n+1}-M_n}
\right. \nonumber \\
\fl \left.\times\left(\frac{1}{M_n - M_{n+1}+\Omega +i (\Gamma_n +
\Gamma_{n+1})}+\frac{1}{M_n - M_{n+1}-\Omega -i (\Gamma_n +
\Gamma_{n+1})}\right)\right. \nonumber \\
\fl
-\left(1+\frac{\Delta^2}{M_nM_{n+1}}\right)\frac{1}{M_{n+1}+M_n}\nonumber
\\ \fl \times\left. \left(\frac{1}{M_n +M_{n+1} +\Omega+ i(\Gamma_n + \Gamma_{n+1})}
+\frac{1}{M_n +M_{n+1} -\Omega- i(\Gamma_n +
\Gamma_{n+1})}\right)\right\}.
\end{eqnarray}
In deriving Eqs.~(\ref{sigma_xx-complex-corrected}),
(\ref{sigma_xy-complex-corrected}) we assumed that the impurity
scattering rate is independent of the energy $\omega$, but kept its
dependence on the Landau level index $n$. This assumption allows us
to eliminate the integration over $\omega$ which is present in
Eq.~(\ref{optical-diagonal}). To preserve the Landau index
dependence, the sum over transitions between neighboring Landau
levels is retained in Eqs.~(\ref{sigma_xx-complex-corrected}),
(\ref{sigma_xy-complex-corrected}).

Based on Eqs.~(\ref{sigma_xx-complex-corrected}),
(\ref{sigma_xy-complex-corrected}) one can easily write down
separate expressions for $\mbox{Re} \, \sigma_{xx}(\Omega)$,
$\mbox{Im} \, \sigma_{xx}(\Omega)$, $\mbox{Re} \,
\sigma_{xy}(\Omega)$, $\mbox{Im} \, \sigma_{xy}(\Omega)$ and verify
that diagonal and off-diagonal conductivities satisfy Kramers-Kronig
relations. A big advantage of
Eqs.~(\ref{sigma_xx-complex-corrected}),
(\ref{sigma_xy-complex-corrected}) is that they are more suitable
for numerical computations and it is sufficient to include only a
few terms in the sum even in  relatively low magnetic field. Also
they are useful for the description of the resonance peaks when the
Landau level index dependence is more important than the energy
dependence of the scattering rate which is included in
Eq.~(\ref{optical-diagonal}).

All figures in this paper were computed for $\Gamma = \mbox{const}$,
so that for calculation of the diagonal conductivity we used
Eq.~(\ref{optical-diagonal}) and checked that it gives almost
identical results to Eq.~(\ref{sigma_xx-complex-corrected}). Since
the corresponding full expression for the optical Hall conductivity
derived in Ref.~\cite{Gusynin2006PRB} is rather difficult for
numerical calculations, in the present paper we compute all results
for the Hall conductivity using
Eq.~(\ref{sigma_xy-complex-corrected}). In the case of  Landau level
index independent width, $\Gamma_n = \mbox{const}$,
Eqs.~(\ref{sigma_xx-complex-corrected}),
(\ref{sigma_xy-complex-corrected})  acquire an even simpler form
\begin{eqnarray}
\label{sigma_xx-complex-corrected1}
\fl \sigma_{xx}(\Omega)=\frac{e^2v_F^2|eB|(\Omega+2i\Gamma)}{\pi
c\,i}  \\
\fl \times \sum_{n=0}^{\infty}
\left\{\left(1-\frac{\Delta^2}{M_nM_{n+1}}\right)\frac{[n_{F}(M_n) -
n_F(M_{n+1})] + [n_F(-M_{n+1}) -
n_F(-M_{n})]}{(M_{n+1}-M_n)^2-(\Omega+2i\Gamma)^2}\frac{1}{M_{n+1}-M_n}
\right. \nonumber \\
\fl
+\left.\left(1+\frac{\Delta^2}{M_nM_{n+1}}\right)\frac{[n_{F}(-M_n)
- n_F(M_{n+1})] + [n_F(-M_{n+1}) -
n_F(M_{n})]}{(M_{n+1}+M_n)^2-(\Omega+2i\Gamma)^2}\frac{1}{M_{n+1}+M_n}\right\}
\nonumber
\end{eqnarray}
and
\begin{eqnarray}
\label{sigma_xy-complex-corrected1}
\fl \sigma_{xy}(\Omega)=-\frac{e^2v_F^2eB}{\pi c}
\sum_{n=0}^{\infty}\left([n_{F}(M_n) - n_F(M_{n+1})] -
[n_F(-M_{n+1}) -
n_F(-M_{n})]\right)   \\
\fl \times \left\{\left(1-\frac{\Delta^2}{M_nM_{n+1}}\right)
\frac{1}{(M_{n+1}-M_n)^2-(\Omega+2i\Gamma)^2} \right. \nonumber \\
\fl \left.
+\left(1+\frac{\Delta^2}{M_nM_{n+1}}\right)\frac{1}{(M_{n+1}+M_n)^2-(\Omega+2i\Gamma)^2}
\right\}. \nonumber
\end{eqnarray}
One can see that the conductivity $\sigma_{xx}(\Omega,\mu)$ is even
function of $\mu$ while $\sigma_{xy}(\Omega,\mu)$ is an odd one.

\subsection{Low field limit of the magneto-optical Lorentzian model}
\label{sec:low-B}

Now we check that Eqs.(\ref{sigma_xx-complex-corrected1}),
(\ref{sigma_xy-complex-corrected1}) reproduce correctly the limit
$B\to0$. Introducing the continuum variable $\omega$ instead of
$M_n$ given by Eq.~(\ref{M_n}) and replacing the sum over $n$ by the
integral, we obtain
\begin{eqnarray}
\label{sigma_xx-lowB} \fl
\sigma_{xx}(\Omega)=-\frac{2ie^2(\Omega+2i\Gamma)}{h}
\left[\frac{1}{(\Omega+2i\Gamma)^2} \int_{\Delta}^{\infty} d \omega
\frac{\omega^2 - \Delta^2}{\omega}\left(\frac{\partial
n_F(\omega)}{\partial \omega}-\frac{\partial n_F(-\omega)}{\partial
\omega} \right) \right.\\
-\left.\int_{\Delta}^{\infty} d\omega \frac{\omega^2 +
\Delta^2}{\omega^2}\frac{n_F(-\omega)-n_F(\omega)}{(\Omega+2i\Gamma)^2-4\omega^2}\right]
\nonumber
\end{eqnarray}
and \begin{eqnarray} \label{sigma_xy-lowB} \fl \sigma_{xy}(\Omega)=
\frac{e^2v_F^2eB}{\pi c}\int_{\Delta}^{\infty} d\omega
\left(\frac{\partial n_F(\omega)}{\partial \omega} +\frac{\partial
n_F(-\omega)}{\partial \omega}\right) \\
 \times
\left[-\frac{\omega^2-\Delta^2}{\omega^2}\frac{1}{(\Omega+2i\Gamma)^2}+
\frac{\omega^2+\Delta^2}{\omega^2}\frac{1}{4\omega^2-(\Omega+2i\Gamma)^2}\right],
\nonumber
\end{eqnarray}
where we restored Planck constant $h = 2 \pi \hbar$ in the overall
prefactor. Here the terms with the factor $(\omega^2
-\Delta^2)/\omega^2$ are intraband and the terms containing the
factor $(\omega^2 + \Delta^2)/\omega^2$ are interband. The
expressions (\ref{sigma_xx-lowB}) and (\ref{sigma_xy-lowB})  are
obtained under the condition $\sqrt{\hbar |eB| v_F^2/c} \ll \Gamma$.
The intraband term of Eq.~(\ref{sigma_xx-lowB}) can be written in
the familiar Drude form
\begin{equation}
\fl \sigma_{xx}^{\mathrm{Drude}} (\Omega) = \frac{2e^2}{h}
\int_{-\infty}^{\infty} d \omega \left(-\frac{\partial
n_F(\omega)}{\partial \omega}\right) \frac{1}{2 \Gamma - i \Omega}
\frac{(\omega^2 - \Delta^2) \theta(\omega^2 - \Delta^2)}{|\omega|}
\end{equation}
which for $\Delta =0$ reduces to Eq.~(2) of
Ref.~\cite{Gusynin2006micro}. The whole expression
(\ref{sigma_xx-lowB}) which includes the interband term, also
reduces to the other limiting cases considered in
Refs.~\cite{Gusynin2006micro,Falkovsky2006}. In particular, an
unusual feature of graphene is that in the high frequency limit the
interband contribution is a constant,
\begin{equation}
\label{diagonal-high-frequency} \sigma_{xx}(\Omega) \simeq \frac{\pi
e^2}{2 h}, \qquad \Omega \gg \mu,\Delta,T.
\end{equation}
Here to rely on the linearized Dirac approximation we assumed that
$\Omega$ is still well below a large band edge.

The real part of the Hall conductivity (\ref{sigma_xy-lowB}) for
$\Omega = \Delta=T=0$ reduces to the expression
\begin{equation}
\sigma_{xy}(\Omega =0) = - \frac{e^2 v_F^2 eB \mbox{sgn} \mu}{4 \pi
c \Gamma^2}, \qquad \sqrt{\hbar |eB| v_F^2/c}\ll \Gamma \ll |\mu|
\end{equation}
which is in agreement with Eq.~(4.3) of Ref.~\cite{Gusynin2006PRB}.
On the other hand, in the high frequency limit
Eq.~(\ref{sigma_xy-lowB}) gives
\begin{equation}
\label{Hall-high-frequency} \sigma_{xy}(\Omega) = \frac{e^2 v_F^2
eB}{\pi  c (\hbar \Omega)^2} \left[\tanh \frac{\mu+ \Delta}{2T} +
\tanh \frac{\mu - \Delta}{2T}\right], \qquad \Omega \to \infty.
\end{equation}
This behavior also follows from
Eq.~(\ref{sigma_xy-complex-corrected}) which is valid in an
arbitrary field $B$. Interestingly, expression
(\ref{Hall-high-frequency}) is sensitive to the relationship between
$|\mu|$ and $\Delta$ and this feature can be used for the gap
detection [see the discussion of Figs.~\ref{fig:4} and \ref{fig:9}
below]. Using Eqs.~(\ref{diagonal-high-frequency}) and
(\ref{Hall-high-frequency}) we obtain the weak field optical Hall
resistivity
\begin{eqnarray}
 \rho_{xy}(\Omega) & = & -
\frac{\sigma_{xy}(\Omega)}{\sigma_{xx}^2(\Omega)+
\sigma_{xy}^2(\Omega)}\nonumber \\ & = & - \frac{16 v_F^2 B}{\pi e c
\Omega^2} \left[\tanh \frac{\mu+ \Delta}{2T} + \tanh \frac{\mu -
\Delta}{2T}\right], \qquad \Omega \gg \mu,\Delta,T.
\end{eqnarray}
Accordingly, the high frequency optical Hall coefficient for
$\Delta=0$ and $T\to 0$
\begin{equation}
R_H(\Omega) = \frac{\rho_{xy}(\Omega)}{B} = -\frac{32v_F^2}{\pi
ec\Omega^2}\,{\rm sgn}(\mu)
\end{equation}
contains the information on the value of the Fermi velocity in
graphene.

Finally, the imaginary part of $\sigma_{xy}(\Omega)$ which follows
from Eq.~(\ref{sigma_xy-lowB}) is given by the expression
\begin{eqnarray}
\label{Imsigma-limit_B0} \fl {\rm
Im}\,\sigma_{xy}(\Omega)=\frac{4e^2v_F^2eB\Gamma\Omega}{\pi
c}\int\limits_\Delta^\infty d\omega \left(\frac{\partial
n_F(\omega)}{\partial
\omega}+\frac{\partial n_F(-\omega)}{\partial \omega} \right) \nonumber \\
\left[ \frac{1}{(\Omega^2-4\Gamma^2)^2+16\Omega^2\Gamma^2}+\frac{1}
{(4y^2-\Omega^2+4\Gamma^2)^2+16\Omega^2\Gamma^2}\right]
\end{eqnarray}
which in high frequency limit becomes
\begin{equation} {\rm
Im}\,\sigma_{xy}(\Omega)\simeq-\frac{4e^2v_F^2eB\Gamma}{\pi
c(\hbar\Omega)^3} \left(\tanh\frac{\mu+\Delta}{2T}+\tanh\frac{\mu -
\Delta}{2T}\right), \qquad \Omega \to \infty.
\end{equation}
The last equation shows that in this limit real part of
$\sigma_{xy}$ given by Eq.~(\ref{Hall-high-frequency}) is the
leading term.

\section{Results for optical conductivity}
\label{sec:cond-results}

\subsection{Diagonal conductivity}
\label{sec:cond-diagonal}

\begin{figure}[h]
\centering{
\includegraphics[width=10cm]{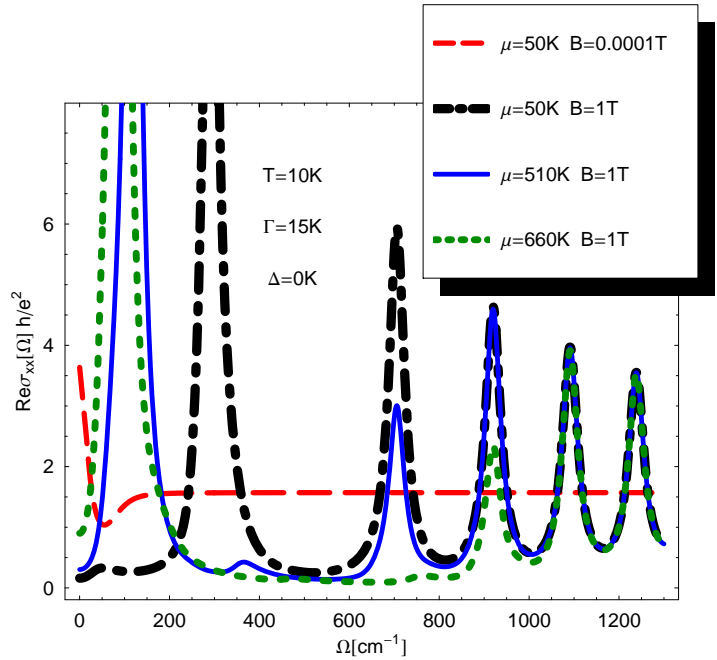}}
\caption{(Colour online) Real part of the longitudinal conductivity,
$\mbox{Re} \, \sigma_{xx}(\Omega)$ in units of $e^2/h$ vs frequency
$\Omega$ in $\mbox{cm}^{-1}$ for temperature $T=10 \mbox{K}$,
scattering rate $\Gamma =15 \mbox{K}$. Long dashed, the chemical
potential $\mu = 50 \mbox{K}$ and the magnetic field $B=10^{-4}
\mbox{T}$, dash-dotted $\mu = 50 \mbox{K}$ and $B=1\mbox{T}$, solid
$\mu = 510 \mbox{K}$ and $B=1\mbox{T}$, short dashed $\mu = 660
\mbox{K}$ and $B=1\mbox{T}$.} \label{fig:1}
\end{figure}

\begin{figure}[h]
\centering{
\includegraphics[width=5.7cm]{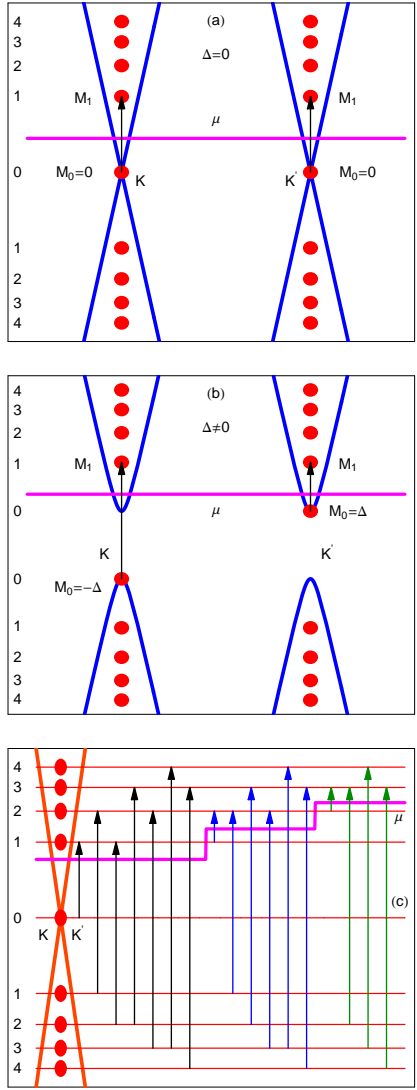}}
\caption{(Colour online) (a) Schematic representation of the two
pairs of Dirac cones with apex at points $\mathbf{K}$ (left) and
$\mathbf{K}^\prime$ (right) in graphene Brillouin zone. The energies
of the Landau levels are shown for index $n=0,\ldots 4$ as solid
(red) circles for both positive and negative Dirac cones. The
transition from $n=0$ to $n=1$ across the chemical potential shown
as a thick horizontal (violet) line are for the case $\Delta=0$. (b)
Same as (a) but now there is a finite excitonic gap $\Delta$. (c)
The pair of cones at points $\mathbf{K}$ and $\mathbf{K}^\prime$ in
the Brillouin zone (see panels (a) and (b) ) are combined. Vertical
arrows show allowed optical transitions between Landau levels for
three values of the chemical potential. At left are the transitions
when $\mu$ is between $n=0$ and $n=1$, in the middle between $n=1$
and $n=2$, and the right between $n=2$ and $n=3$. The first line in
the left is different in that it is both inter (between two separate
cones) and intra (within a given cone). The first line in all other
series in intra and the shortest inter appears only once, while all
others appear twice. Note that from
Eq.~(\ref{sigma_xx-complex-corrected1}) $\sigma_{xx}(\Omega,\mu)$ is
even in $\mu$, while from Eq.~(\ref{sigma_xy-complex-corrected1})
$\sigma_{xy}(\Omega,\mu)$ is odd in $\mu$. } \label{fig:2}
\end{figure}

In Fig.~\ref{fig:1} we show the results based on a numerical
evaluation of the full equation (\ref{optical-diagonal}) for the
real part of the longitudinal conductivity $\mbox{Re} \,
\sigma_{xx}(\Omega)$ in units $e^2/ h$ as a function of frequency in
units of $\mbox{cm}^{-1}$. Except for the long dashed (red) curve
which was obtained in the limit of vanishing external magnetic field
(namely $B=10^{-4} \mbox{T}$) and is included for comparison, the
other three curves are for $B=1 \mbox{T}$. They differ in value of
the chemical potential $\mu$. In all cases the temperature $T=10
\mbox{K}$, the impurity scattering $\Gamma = 15 \mbox{K}$, and the
excitonic gap $\Delta=0$. For reference in scrutinizing the curves,
the frequency of the $n=1$ Landau level, $\Omega_1 =M_1(\Delta=0)=
294 \mbox{cm}^{-1} [423 \,\mbox{K}]$, $\Omega_2 =
M_2(\Delta=0)=415.8 \mbox{cm}^{-1} [598\, \mbox{K}]$ and $\Omega_3 =
M_3(\Delta=0)=509.2 \mbox{cm}^{-1} [733\, \mbox{K}]$, so that for
the dash-dotted (black) curve $\mu$ falls below the energy of the
$n=1$ level (see left side of Fig.~\ref{fig:2}~(c)) \footnote{We
recall that the conversion rule from the frequency in
$\mbox{cm}^{-1}$ to the energy in Kelvins is $\Omega [\mbox{K}] =
1.4387 \mbox{K} \cdot \mbox{cm} \Omega[\mbox{cm}^{-1}]$.}, for the
solid (blue) curve it falls between $n=1$ and $n=2$ levels (see
middle of Fig.~\ref{fig:2}~(c)), and for the short dashed (green) it
is between $n=2$ and $n=3$ (right side of Fig.~\ref{fig:2}~(c)). The
energies of the peaks are $M_1$, $M_1 + M_2$, $M_2+M_3$, etc. and
$M_2 - M_1$, $M_3 -M_2$, etc. Note that the position and the
intensity of the last two peaks in Fig.~\ref{fig:1} (largest
frequency $\Omega$) remains the same for all chosen values of the
chemical potential $\mu$.
When $\mu$ falls between $M_2$ and $M_3$ [short dashed (green)
curve] the intensity in the third last peak has dropped to half the
value it has in solid (blue) curve, while the fourth last peak  has
merged into a low intensity background, as has the fifth last, which
is seen only in the dash-dotted curve. Also a new peak has appeared
at $M_3-M_2$ which was not present in the dash-dotted curve.
Similarly for $\mu$ between $M_1$ and $M_2$ [solid (blue) curve] the
intensity of the fourth highest energy peak has dropped to half the
intensity it has in the dash-dotted curve and the peak at $M_1$ is
entirely missing, having merged into the low intensity background. A
new peak has appeared at $M_2 -M_1$. Finally when $\mu$ is below
$M_1$ (dash-dotted curve) there is no peak below $M_1$ and the line
at $M_1$ always has full intensity. {\em Whatever the value of $\mu$
this line will never be seen to half its intensity and this is the
hallmark of the Dirac nature of the quasiparticles.} A schematic
which helps us understand the behaviour of the absorption lines that
we have just described is given in Fig.~\ref{fig:2}~(c).
On the left of the figure we show the energies of the Landau levels
$E_n= \pm M_n$ of Eq.~(\ref{M_n}) as solid (red) dots along with
their quantum numbers $n=0,1,2,\ldots$. The Dirac cones which come
in pairs, with the positive and negative energies and exist at
$\mathbf{K}$ and $\mathbf{K}^\prime$ points in graphene Brillouin
zone are also shown. The three values of chemical potential
considered in Fig.~\ref{fig:2}~(c) are shown in heavy solid (violet)
horizontal lines. The possible optical transitions in each case are
indicated as vertical arrows and connect levels $n$ to $n\pm 1$
only. Moving from left to right we see  first a single transition
from $E_0$ to $E_1 =M_1$, then a pair of interband from $E_1=-M_1$
to $E_2=M_2$ and $E_2 =-M_2$ to $E_1=M_1$ followed by another pair
from $E_2 = -M_2$ to $E_3 = M_3$ and $E_3 = -M_3$ to $E_2 =M_2$ etc.
For the middle set of lines the first is an intraband transition
from $E_1=M_1$ to $E_2 = M_2$ followed by a single interband from
$E_1 = -M_1$ to $E_2 = M_2$ and then a pair from $E_2 = -M_2$ and
$E_3 = M_3$ and $E_3 = -M_3$ to $E_2 =M_2$ etc. Finally in the set
of transitions at the right of the figure there is an intraband from
$E_2 = M_2$ to $E_3 = M_3$ followed by a single interband from $E_2
= - M_2$ to $E_3 = M_3$ and a pair from $E_3 = -M_3$ to $E_4 = M_4$
and $E_4 = -M_4$ to $E_3 = M_3$ etc. This is precisely the pattern
we have seen in Fig.~\ref{fig:1}. We note one more aspect of the
anomalous line at the far left of the figure for the transition form
$n=0$ to $n=1$ Landau level. It is the only line which cannot be
unambiguously assigned to inter or intraband because the sate at
$n=0$ falls at the apex of the Dirac cones, where positive and
negative energy cones meet and hence they share this state equally.
This is further illustrated in Figs.~\ref{fig:2}~(a) and (b), where
two sets of cones at $\mathbf{K}$ and $\mathbf{K}^\prime$ are shown
separately for the case of $\Delta=0$ in frame (a) and finite
excitonic gap in frame (b). In this second case we see clearly that
points $\mathbf{K}$ and $\mathbf{K}^\prime$ react differently under
a finite magnetic field. For the cone on the left $n=0$ level has
moved to energy $E_0 = - \Delta$ and for the cone on the right it
has moved to $E_0 = \Delta$. Note that for the value of chemical
potential shown as a solid horizontal (violet) line, the
$E_0=-\Delta$ to $E_1$ transition (vertical arrow) on the left is
now definitely interband and the $E_0 = \Delta$ to $E_1$ on the
right is intraband. The ambiguity in designation of the $n=0$ level
present in the top frame (a) of Fig.~\ref{fig:2} is lifted when the
gap becomes finite.

The rather complicated pattern of behaviour just described can be
understood simply from the mathematics of the previous section in
the limit $\Gamma\to0$ and $T \to 0$. Taking $\Delta=0$, $\mu\geq
0$, and $\Omega \geq0$ one obtains from Eq.~(\ref{Lorentz-full})
\begin{eqnarray}
\label{sigma-T=G=0}
\fl \mbox{Re} \, \sigma_{xx}(\Omega) = \frac{e^2}{h}M_1^2
\frac{\pi}{2} \sum_{n=0}^{\infty} \left\{[2-n_F(M_n)-n_F(M_{n+1})]
\frac{\delta(M_n+M_{n+1}-\Omega)}{M_n+ M_{n+1}} \right. \nonumber \\
 \left. \lo+ [n_F(M_n)-n_F(M_{n+1})]
\frac{\delta(M_n-M_{n+1}+\Omega)}{M_{n+1}-M_n} \right\}.
\end{eqnarray}
For $\mu \in ]M_N,M_{N+1}[$  the $T=0$ thermal factor
\begin{equation}
\label{intra} 2-n_F(M_n)-n_F(M_{n+1}) =  \left\{
\begin{array}{c c c}
0 & \mbox{for} & n< N, \\
1 & \mbox{for} &  n=N, \\
2 & \mbox{for} & n> N,
\end{array} \right.
\end{equation}
while
\begin{equation}
\label{inter} n_F(M_n)-n_F(M_{n+1})=  \left\{
\begin{array}{c c c}
1 & \mbox{for} & n= N, \\
0 & \mbox{for} &  n\neq N.
\end{array} \right.
\end{equation}
The line in $\mbox{Re} \, \sigma_{xx}(\Omega)$ with the frequency
$M_n + M_{n+1}$ occurs only for $n\geq N$ with the others missing.
Further the line with $n=N$ has half of the weight it would have if
it occurred in another case, namely, $1/(M_n+M_{n+1})$, while lines
for $n>N$ have full weight $2/(M_n+M_{n+1})$. Note that the lines
with the frequency equal to the difference in Landau level energies
have weight $1/(M_{n+1} -M_n)$. These lines are always present
except for the case when $\mu$ falls below $M_1$ when the single
line at $M_1$ has weight $2/M_1$. In summary, we have seen in the
above discussion that as $\mu$ moves through higher and higher
values of $M_N$ the lines below $N$ disappear into the background
with a new line appearing at $M_{N+1} -M_{N}$. Further the line at
$n=N$ loses half its intensity while the others remain unaltered
except for the special case when chemical potential falls below the
Landau level $M_1$.

The pattern would be quite different if instead of $M_n \sim
\sqrt{n}$ the Landau level quantization was Schr\"odinger-like $M_n=
\omega_{c}(n+1/2)$ with $\omega_c$ being the cyclotron frequency. In
this case the position of the line corresponding to the difference
in Landau level energies (intraband) never shifts in energy.
Further, the lines corresponding to the sum of the Landau level
energies (interband) fall at regular energies intervals
$2\omega_c(n+1)$, namely $2\omega_c$, $4 \omega_c$, $6\omega_c$,
\ldots. Furthermore as $\mu$ increases through the energies of the
various Landau levels all lines half their intensity before fading
into the background and for $\mu$ below the first level there is no
line at $\omega_c$. There is also no ambiguity about whether a line
is intra or interband as no level is shared between upper and lower
cone.

In field effect devices the chemical potential in a graphene sheet
can be changed by applying a gate voltage and this may be an ideal
way to observe the effects just described. However, there is an
alternative way to see the same effects. For any fixed value of
chemical potential, the external magnetic field can be selected in
such a way that $\mu$ falls below the first Landau level or in
between $n=1$ and $n=2$ etc. For appropriate choices of $B$
\cite{Gusynin2006optical-short} the curves for $\mbox{Re} \,
\sigma_{xx}(\Omega)$ can be made to behave exactly as in
Fig.~\ref{fig:1}. To see this it is important that $\Omega$ be
divided by $\sqrt{\hbar e B v_F^2/c}$, so that the lines remain
fixed in normalized frequency and the vertical scale be divided by
the same factor and multiplied by $\Gamma$ to  keep the dimensions
the same. When this is done, the same pattern as seen in
Fig.~\ref{fig:1} emerges for this configuration corresponding to
fixed $\mu$ with several well chosen values of $B$.

We have found that the curves of Fig.~\ref{fig:1} change very little
as the chemical potential is varied within the limited range $M_N$
to $M_{N+1}$. We have also verified that for the parameters used
here $T=10 \mbox{K}$, $\Gamma=15 \mbox{K}$ the crossover from half
intensity in the line $M_{N}+ M_{N+1}$ to near zero, i.e. merging
into the background occurs rather abruptly in an energy range set by
$T$ and/or $\Gamma$.
\begin{figure}[h]
\centering{
\includegraphics[width=10cm]{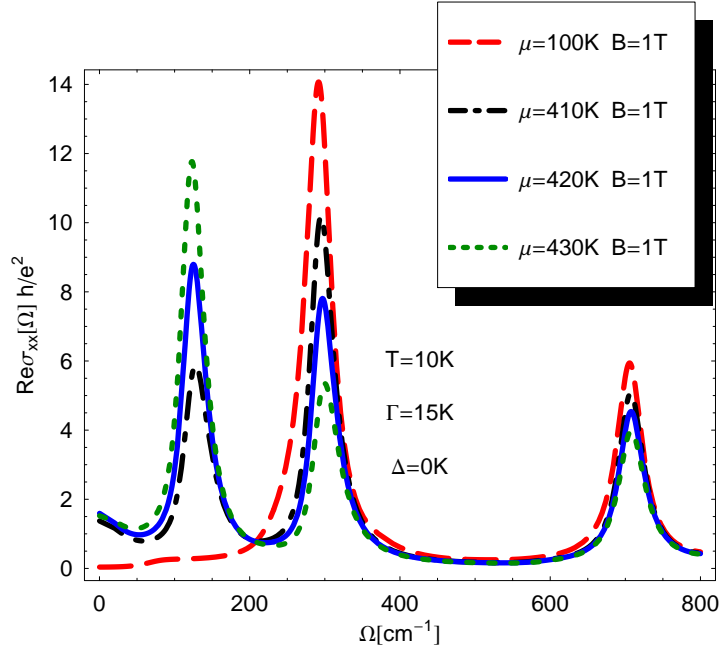}}
\caption{(Colour online) Real part of the longitudinal conductivity,
$\mbox{Re} \, \sigma_{xx}(\Omega)$ in units of $e^2/h$ vs frequency
$\Omega$ in $\mbox{cm}^{-1}$ for temperature $T=10 \mbox{K}$,
scattering rate $\Gamma =15 \mbox{K}$ and magnetic field $B=1
\mbox{T}$ for 4 values of chemical potential. Long dashed $\mu = 100
\mbox{K}$, dash-dotted $\mu = 410 \mbox{K}$, solid  $\mu = 420
\mbox{K}$, short dashed $\mu = 430 \mbox{K}$.} \label{fig:3}
\end{figure}
This is illustrated in Fig.~\ref{fig:3} where we consider how the
complete disappearance of the peak at  $\Omega = \Omega_1 = 294
\mbox{cm}^{-1}$ ($420 \mbox{K}$) and the depletion of the next
higher peak at $710 \mbox{cm}^{-1}$ towards half its initial value
proceeds as the chemical potential $\mu$ crosses through the energy
of the $n=1$ Landau level [$\Omega_1 =M_{1}(\Delta=0)$]. The long
dashed curve (red) is for $\mu=100 \mbox{K}$ chosen to be well away
from the crossover point of $420 \mbox{K}$ and is shown for
comparison. The dash-dotted curve (black) is for $\mu=410 \mbox{K}$
slightly below the crossover point, the solid curve (blue) is for
$\mu=420 \mbox{K}$ just at the crossover energy and the short dashed
curve (green) is for $\mu=430 \mbox{K}$ which spans $\pm 10 \,
\mbox{K}$ on either side of $\Omega_1$ which is much less than the
level width of $2\Gamma = 30\, \mbox{K}$. As $\mu$ increases through
$410 \mbox{K}$ we note the growth of the peak at $\Omega = 122
\mbox{cm}^{-1}$, the depletion towards zero value of the peak at
$294 \mbox{cm}^{-1}$ and towards half its initial value of the next
higher peak at $710 \mbox{cm}^{-1}$. The complete transfer of
spectral weight between the various peaks is completed for a rather
small region of chemical potential variation about $\Omega_1 = 294
\mbox{cm}^{-1}$ ($420 \mbox{K}$) of order $30 \, \mbox{K}$ (not
shown in the figure). After the crossover is complete, the pattern
of the spectral weight distribution will remain unchanged until the
chemical potential becomes close to the energy of the next Landau
level. We would like to stress that although Fig.~\ref{fig:3} is
plotted using the full expression (\ref{optical-diagonal}), we have
verified that the results obtained using Eqs.~(\ref{Lorentz-full})
and (\ref{sigma_xx-complex-corrected}) are practically identical.

Returning to Eqs.~(\ref{sigma-T=G=0}) to (\ref{inter}) we note that
the optical spectral weight under a given inter- or intraband line
varies as the  square root of $B$ (see Eq.~(\ref{weight-B-final})
and the discussion associated with it). This has been verified in
the recent experiment of Sadowski {\em et al.} \cite{Sadowski2006}
on ultrathin epitaxial graphite samples
\cite{Berger2004JPCB,Berger2006Science}.

\begin{figure}[h]
\centering{
\includegraphics[width=10cm]{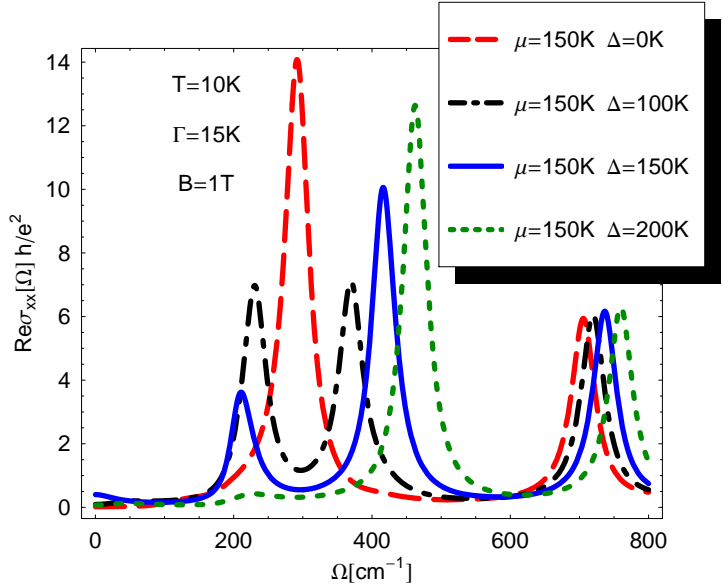}}
\caption{(Colour online) Real part of the longitudinal conductivity,
$\mbox{Re} \, \sigma_{xx}(\Omega)$ in units of $e^2/h$ vs frequency
$\Omega$ in $\mbox{cm}^{-1}$ for temperature $T=10 \mbox{K}$,
$\Gamma =15 \mbox{K}$, $B=1 \mbox{T}$ and chemical potential $\mu =
150 \mbox{K}$ for 4 values of the excitonic gap $\Delta$. Long
dashed $\Delta = 0 \mbox{K}$, dash-dotted $\Delta = 100 \mbox{K}$,
solid $\Delta = 150 \mbox{K}$, short dashed $\Delta = 200
\mbox{K}$.} \label{fig:4}
\end{figure}
The effect of an excitonic gap $\Delta$ on the optical spectral
weight distribution in a one Tesla magnetic field is illustrated in
Fig.~\ref{fig:4}. Here as in the previous figures $\Gamma=15
\mbox{K}$ and $T=10 \mbox{K}$. The chemical potential is set at $150
\mbox{K}$ in all cases. The long dashed (red),  dash-dotted (black),
solid (blue) and short dashed (green) curves are for $\Delta=0, 100
\mbox{K}$, $150 \mbox{K}$ and $200 \mbox{K}$ respectively. The curve
for $\Delta=0$ is for reference. We note that for finite $\Delta=100
\mbox{K}$ the line at $\Omega_1 = M_1(\Delta=0)=294 \mbox{cm}^{-1}$
splits into two peaks [dash-dotted (black) curve]. The lower peak is
at energy $M_1(\Delta) - M_0(\Delta)=\sqrt{\Omega_1^2 +
\Delta^2}-\Delta$, while the upper peak is at $M_1(\Delta) +
M_0(\Delta)=\sqrt{\Omega_1^2 + \Delta^2}+\Delta$. These two
transitions for that value of chemical potential are illustrated in
Fig.~\ref{fig:2} middle frame~(b) (see arrows). Additional
transitions not shown in this frame  can, of course, occur but these
will have higher energy. For $\Delta=150 \mbox{K}=\mu$ the $n=0$
state shown on the right hand cone of Fig.~\ref{fig:2}~(b) is
occupied with a probability $1/2$ at $T=0$, $\Gamma=0$ and a
transition from $n=1$ lower cone to $n=0$ upper cone is possible as
is from $n=0$ upper cone to $n=1$ of the same cone and each must be
weighted by a factor $1/2$. The first has energy $\sqrt{\Omega_1^2 +
\Delta^2}+\Delta$, while the second has energy $\sqrt{\Omega_1^2 +
\Delta^2}-\Delta$. There is an additional transition coming from the
second cone on the left hand side of the figure of energy
$\sqrt{\Omega_1^2 + \Delta^2}+\Delta$. Thus the spectral intensity
of the lower energy line in the solid curve (blue) is lower than
that of the higher energy line by a factor of $3$. Finally for
$\Delta=200 \mbox{K}$ $(\Delta>\mu)$ the energy of the lowest
transition possible in both left and right side cones are the same
equal to $\sqrt{\Omega_1^2 + \Delta^2}+\Delta$ and so there is only
one line in the dashed (green) curve. This pattern of behaviour
should allow one to measure the occurrence of an excitonic gap
catalyzed by the magnetic field. We note that the splitted peak in
Fig.~\ref{fig:4} with finite $\Delta$ acquire a slight asymmetry
because of the factor $1 \pm \Delta^2/M_n M_{n+1}$ of
Eq.~(\ref{Lorentz-full}) which is different for inter and intraband
transitions.

\begin{figure}[h]
\centering{
\includegraphics[width=10cm]{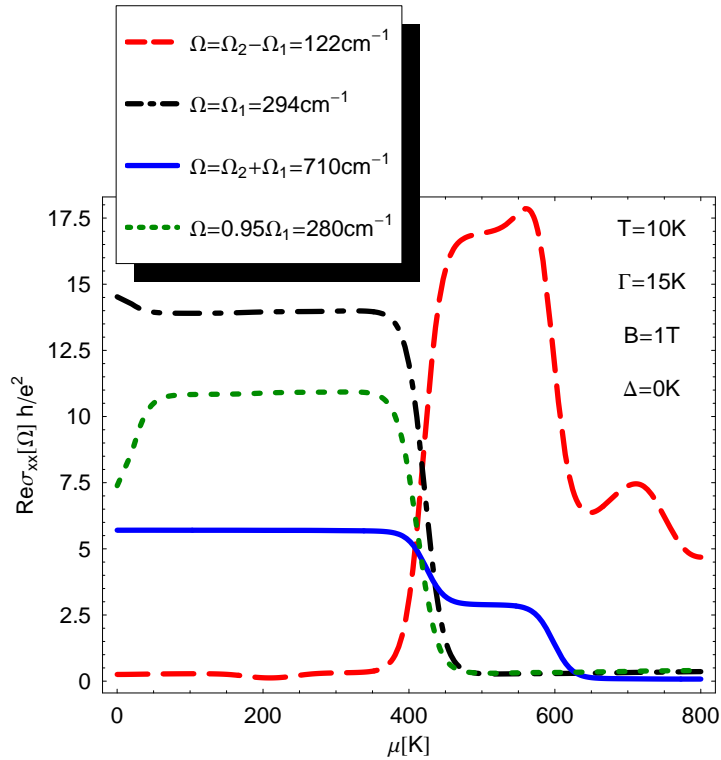}}
\caption{(Colour online) Real part of the longitudinal conductivity,
$\mbox{Re} \, \sigma_{xx}(\Omega)$ in units of $e^2/h$ vs the
chemical potential $\mu$ in $\mbox{K}$ for  $T=10 \mbox{K}$, $\Gamma
=15 \mbox{K}$, $B=1 \mbox{T}$ and $\Delta=0$. Four frequencies
$\Omega$ are considered, long dashed $\Omega = 122 \mbox{cm}^{-1}$,
dash-dotted $\Omega = 294 \mbox{cm}^{-1}$, solid $\Omega = 710
\mbox{cm}^{-1}$, short dashed $\Omega = 0.95 \Omega_1=
280\mbox{cm}^{-1}$. The first three frequencies correspond to
$\Omega=\Omega_2 -\Omega_1$, $\Omega_1$ and $\Omega_2+\Omega_1$,
respectively, for the parameters used.} \label{fig:5}
\end{figure}

So far we have shown the results for fixed values of chemical
potential and magnetic field as a function of photon energy. It is
also of interest to fix the photon frequency and sweep either
chemical potential ($\mu$) or field ($B$) \cite{Basov-private}. We
begin with the first of these two possibilities and this is
illustrated in Fig.~\ref{fig:5}.
Four well chosen frequencies are selected, viz. $\Omega=\Omega_2 -
\Omega_1 = 122 \mbox{cm}^{-1}$ long dashed curve (red), $\Omega =
\Omega_1 = 294 \mbox{cm}^{-1}$ dash-dotted curve (black),
$\Omega=\Omega_2 + \Omega_1 = 710 \mbox{cm}^{-1}$ solid curve
(blue), and $\Omega=0.95 \Omega_1 = 280 \mbox{cm}^{-1}$ short dashed
curve (green). Taking these in order, we note that the long dashed
curve (red) is near zero until $\mu$ reaches the value of $420
\mbox{K}$ at which point it increases rapidly reaching a plateau at
$\mbox{Re} \, \sigma_{xx}(\Omega) \approx 17.5 e^2/h$ [the height
being set by the height of the peak in the solid (blue) curve in
Fig.~\ref{fig:1}] after which it drops rapidly as $\mu$ goes through
$595 \mbox{K}$. It does not drop down all the way to zero however as
it continues to sample parts of the peaks at the difference in
frequencies $M_{n+1}-M_n$ until these move below the sampling
optical frequency set at $122 \mbox{cm}^{-1}$. The next optical
frequency chosen at $\Omega=\Omega_1$ short-long dashed (black)
curve begins by sampling the equivalent curve in Fig.~\ref{fig:1}
where the peak has height $\sim 14 e^2/h$. However when $\mu$
crosses the energy of the first Landau level at $420 \mbox{K}$ this
peak disappears and the curve drops to zero. No other level crosses
this frequency again. The short dashed curve is for $\Omega=0.95
\Omega_1$. It shows very much the same behaviour as the previous
case, but its plateau height is a little smaller because the peak
(dash-dotted curve of Fig.~\ref{fig:1}) is sampled not at its
center, but rather slightly below its maximum value. The final solid
(blue) curve is for $\Omega=\Omega_2 + \Omega_1$. In this case a
first plateau is seen at small $\mu$ with height $\sim 5.5 e^2/h$
which is the height of the second peak in the dash-dotted (black)
curve of Fig.~\ref{fig:1}. It drops to half of this value as
$\Omega$ crosses $420 \mbox{K}$ and then near zero as $\Omega$
crosses $595 \mbox{K}$. All features of these curves can be traced
to the corresponding behaviour of the curves of Fig.~\ref{fig:1}. An
additional feature of these curves in now described. Note that as
the chemical potential gets small the black (dash-dotted) curve
increases slightly, while the green (dotted) curve drops. As we have
described in connection with Fig.~\ref{fig:3} the line shapes do
depend on the value of $\mu$ if it falls within order $\Gamma$
and/or $T$ of a Landau level energy $M_n$. For $\mu$ near zero, the
first peak height in the black (dash-dotted) curve of
Fig.~\ref{fig:1} increases by $\sim 5 \%$ and its width is also
slightly narrowed. This leads to a slight increase in peak height at
$\Omega = \Omega_1 = 294 \mbox{cm}^{-1}$ monitored in the black
(dash-dotted) curve of Fig.~\ref{fig:5} and a reduction in the green
(dotted) curve which monitors the height of the curve slightly off
the peak at $\Omega = 0.95 \Omega_1 = 280 \mbox{cm}^{-1}$.

Another set of useful curves when considering possible experimental
configurations is to fix the frequency of the light as well as the
chemical potential and sweep the magnetic field.
\begin{figure}[h]
\centering{
\includegraphics[width=10cm]{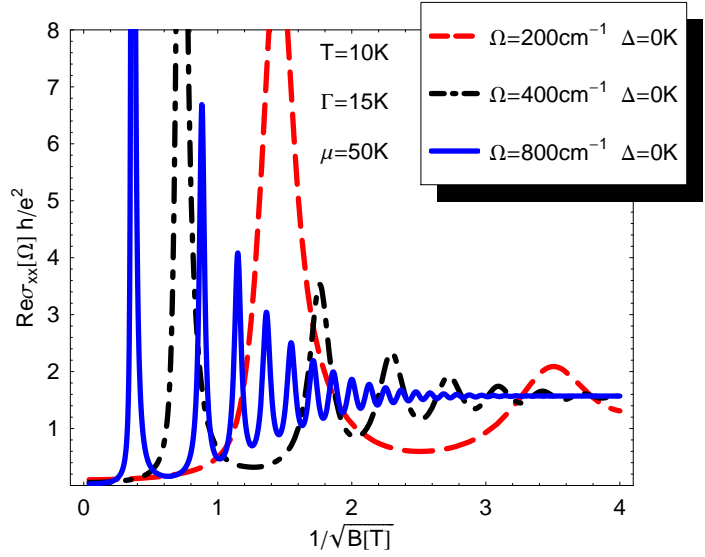}}
\caption{(Colour online) Real part of the longitudinal conductivity,
$\mbox{Re} \, \sigma_{xx}(\Omega)$ at three frequencies, long dashed
$200 \mbox{cm}^{-1}$, dash-dotted $400 \mbox{cm}^{-1}$, solid $800
\mbox{cm}^{-1}$ as a function of the inverse of the square root of
the magnetic field, $1/\sqrt{B}$ with $B$ in $\mbox{T}$. The other
parameters are $T=10 \mbox{K}$, $\Gamma =15 \mbox{K}$ and $\mu=50
\mbox{K}$.} \label{fig:6}
\end{figure}
Results are shown in Fig.~\ref{fig:6} for three cases $\Omega =200
\mbox{cm}^{-1}$ long dashed (red) curve, dash-dotted (black) curve
for $\Omega=400 \mbox{cm}^{-1}$ and solid (blue) for $\Omega=800
\mbox{cm}^{-1}$. In all cases $T=10 \mbox{K}$, $\Gamma=15 \mbox{K}$,
$\mu=50 \mbox{K}$ and the excitonic gap $\Delta$ is taken to be
zero. On the horizontal axis we plotted $1/\sqrt{B(\mbox{T})}$, the
inverse square root of the magnetic field in Tesla, so that fields
below one tesla fall above one on this scale. The pattern of
oscillations is perhaps more complex than in previous curves, but
can easily be traced out from a knowledge of such curves for
$\mbox{Re} \, \sigma_{xx}(\Omega)$ at different values of $B$. As an
example consider the case $\Omega =200 \mbox{cm}^{-1}$. The value of
$\mbox{Re}\,\sigma_{xx}(\Omega) h/e^2$ at $1/\sqrt{B(\mbox{T})}=1$
is just that of the dash-dotted (black) curve of Fig.~\ref{fig:1}
read at $\Omega =200 \mbox{cm}^{-1}$. As $B$ is decreased from its
value $1 \mbox{T}$, the first peak in dash-dotted curve in
Fig.~\ref{fig:1} will move to lower frequency and so increases
significantly the value of longitudinal conductivity at $\Omega =200
\mbox{cm}^{-1}$. Its peak will cross this reference energy for
$1/\sqrt{B(\mbox{T})} \simeq 1.47$ and this produces the first peak
in the long dashed (red) curve of Fig.~\ref{fig:6}. As $B$ is
decreased further below one Tesla, the second peak in the
dash-dotted curve of  Fig.~\ref{fig:1} will also move through
$\Omega =200 \mbox{cm}^{-1}$. This occurs for $1/\sqrt{B(\mbox{T})}
\simeq 3.5$, where a second lower intensity peak is seen in the long
dashed (red) curve of Fig.~\ref{fig:6}. The other curves of
Fig.~\ref{fig:6} can  be traced out from similar considerations
based on Fig.~\ref{fig:1} with attention paid to the evolution of
these curves with changing value of $B$.

\subsection{Spectral weight}
\label{sec:weight}

An interesting quantity to consider is the optical spectral weight
that falls between $\Omega=0$ and $\Omega=\Omega_m$ with $\Omega_m$
a variable upper limit in the integral
\begin{equation}\label{weight-def}
W(\Omega_m) = \int_{0}^{\Omega_m} d \Omega \mbox{Re} \,
\sigma_{xx}(\Omega).
\end{equation}
For the case $T= \Gamma = \Delta = B=0$ we have
\cite{Gusynin2006micro}
\begin{equation} \label{B=0.cond-intra-inter-T=0}
\sigma_{xx}(\Omega)= \frac{\pi
e^2N_f}{h}|\mu|\delta(\Omega)+\frac{\pi
e^2N_f}{4h}\theta\left(\frac{|\Omega|}{2} -|\mu|\right).
\end{equation}
With $N_f=2$ this leads to
\begin{equation}\label{weight-B=0}
\fl W(\Omega_m) = \frac{e^2}{h} (\pi |\mu| + \theta(\Omega_m/2 -
|\mu|) \pi (\Omega_m/2 - |\mu|)) \simeq \frac{e^2}{h} \frac{\pi
\Omega_m}{2} \quad \mbox{for} \quad \Omega_m \gg |\mu|.
\end{equation}
For finite $B$, assuming for simplicity $M_0 < \mu < M_1$  we get
from Eq.~(\ref{sigma-T=G=0})
\begin{equation}\label{sigma-T=0}
\sigma_{xx}(\Omega) = \frac{e^2 N_f v_F^2|eB|}{2 c}
\sum_{n=0}^{\infty} \frac{\delta(M_n + M_{n+1} -
\Omega)}{M_n+M_{n+1}}.
\end{equation}
Accordingly for $\Delta=0$
\begin{eqnarray}
\label{weight-B} \fl W(\Omega_m)=\frac{e^2 N_f v_F^2|eB|}{2 c}
\sum_{n=0}^{\infty}\frac{\theta(\Omega_m - M_n - M_{n+1})}{M_n +
M_{n+1}} \nonumber\\
\lo= \frac{\pi e^2}{h} \sqrt{2 \hbar |eB|
v_F^2/c} \sum_{n=0}^{N}\frac{1}{\sqrt{n}+ \sqrt{n+1}}.
\end{eqnarray}
For  large  values $\Omega_m$ the maximum $N$ that contributes to
(\ref{weight-B}) can be estimated from $\Omega_m = M_N + M_{N+1}
\simeq 2 M_N$ and so $N = \frac{\Omega_m^2}{8 \hbar |eB| v_F^2/c}$.
But
\begin{equation}
\sum_{n=0}^{N}\frac{1}{\sqrt{n}+ \sqrt{n+1}} = \sqrt{N+1} \approx
\sqrt{N} \quad \mbox{for} \quad N \gg 1,
\end{equation}
and so
\begin{equation}
\label{weight-B-final} W(\Omega_m) \simeq \frac{e^2}{h} \frac{\pi
\Omega_m}{2}
\end{equation}
which agrees with Eq.~(\ref{weight-B=0}) as we would expect. We note
that as we have stated, the area under each line goes as $\sqrt{B}$,
but the sum over the lines up to $\Omega_m$ gives another factor
$\sim 1/\sqrt{B}$ which means that $W(\Omega_m)$ is independent of
$B$.
\begin{figure}[h]
\centering{
\includegraphics[width=10cm]{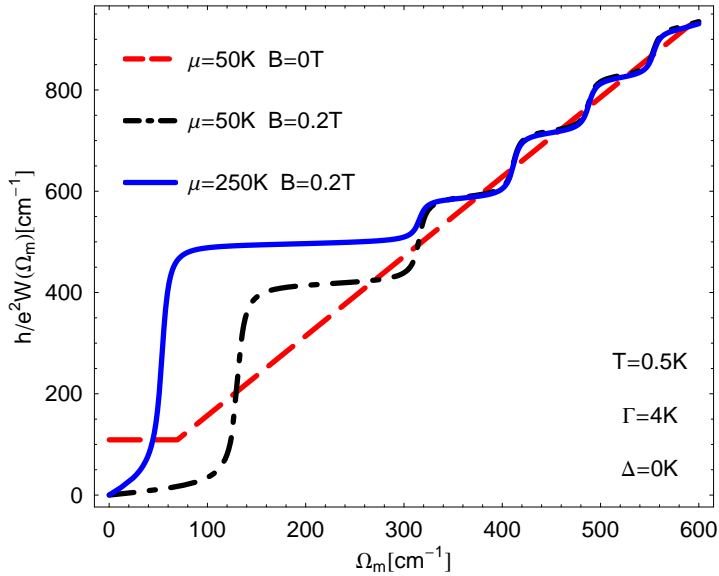}}
\caption{(Colour online) Variation of the optical sum $W(\Omega_m)$
(multiplied by $h/e^2$) in $\mbox{cm}^{-1}$ as a function of
$\Omega_m$ in $\mbox{cm}^{-1}$ for three different cases. Long
dashed $\mu = 50 \mbox{K}$, $B=0 \mbox{T}$ and $T=\Gamma=0 \mbox{K}$
(see Eq.~(\ref{weight-B=0})); dash-dotted $\mu =50 \mbox{K}$, $B=0.2
\mbox{T}$, $T=0.5 \mbox{K}$ and $\Gamma=4 \mbox{K}$; solid $\mu =250
\mbox{K}$, $B=0.2 \mbox{T}$, $T=0.5 \mbox{K}$ and $\Gamma=4
\mbox{K}$. } \label{fig:7}
\end{figure}
In Fig.~\ref{fig:7} we show numerical results for
$(h/e^2)W(\Omega_m)$ in units of $\mbox{cm}^{-1}$ as a function of
$\Omega_m$ is $\mbox{cm}^{-1}$. The long dashed (red) curve is
computed on the base of the first equality in Eq.~(\ref{weight-B=0})
for $\mu=50 \mbox{K}$, $T=0 \mbox{K}$, and $B=0 \mbox{T}$ and is
given for comparison with the two other curves. These curves are
obtained by numerical integration of Eq.~(\ref{weight-def}) with
$\mbox{Re} \, \sigma_{xx}(\Omega)$ given by Eq.~(\ref{Lorentz-full})
for the case $T=0.5 \mbox{K}$, $\Gamma= 4 \mbox{K}$, $\Delta=0 $ and
finite $B=0.2 \mbox{T}$. The dash-dotted (black) curve is for
chemical potential $\mu=50 \mbox{K}$, while solid (blue) is for
$\mu=250 \mbox{K}$. For the first case $\mu$ is below $M_1$ at
$131.5 \mbox{cm}^{-1}$ and for the second  is above. At small values
of $\Omega_m$ both curves start at zero and remain small until
$\Omega_m$ goes through the first Landau peak in Fig.~\ref{fig:1} at
which point it rises sharply and subsequently exhibits a plateau
before showing a next sharp rise. For the dash-dotted curve (black)
the first rise is at $131.5 \, \mbox{cm}^{-1}$ (interband line), but
for the solid (blue) curve it is at the intraband line ($54.6\,
\mbox{cm}^{-1}$). As $\Omega_m$ increases to include several peaks
in $\sigma_{xx}(\Omega)$ the curves start to follow more closely the
$B=0 \mbox{T}$ result [long dashed (red) curve]. This behaviour can
be understood better from a consideration of Eq.~(\ref{weight-B}).
The magnetic field mainly readjusts the available optical spectral
weight among the Landau levels below $\Omega_m$.

To be specific we have considered explicitly in this section only
the case $M_0< \mu < M_1$. For $\mu \in ]M_{N},M_{N+1}[$  we can
show that the missing spectral weight in the lines $n\leq N$ is all
to be found in the single intraband line at $M_{N+1}-M_N$. From
Eq.~(\ref{sigma-T=G=0}) and noting Eqs.~(\ref{intra}) and
(\ref{inter}) as $T \to 0$, the optical spectral weight lost in
units of $(e^2/h) M_1 (\pi/2)$ is
\begin{equation}
\label{weight-conserve} \sum_{n=0}^{N-1}
\frac{2}{\sqrt{n+1}+\sqrt{n}} + \frac{1}{\sqrt{N+1}+\sqrt{N}} \equiv
\frac{1}{\sqrt{N+1}-\sqrt{N}} .
\end{equation}
The first term in the left hand side is the spectral weight from all
the lines that have completely disappeared from $n=0$ to $N-1$. The
second term is from the reduction in intensity by a factor of $1/2$
of the line at $n=N$. The quantity in the right hand side is the
optical weight of the intraband line which has picked up all of the
lost intensity.

\subsection{Hall conductivity}
\label{sec:cond-Hall}

Next we consider the absorptive part of the transverse Hall
conductivity. The calculations are based on the  equation
(\ref{sigma_xy-complex-corrected}).
\begin{figure}[h]
\centering{
\includegraphics[width=10cm]{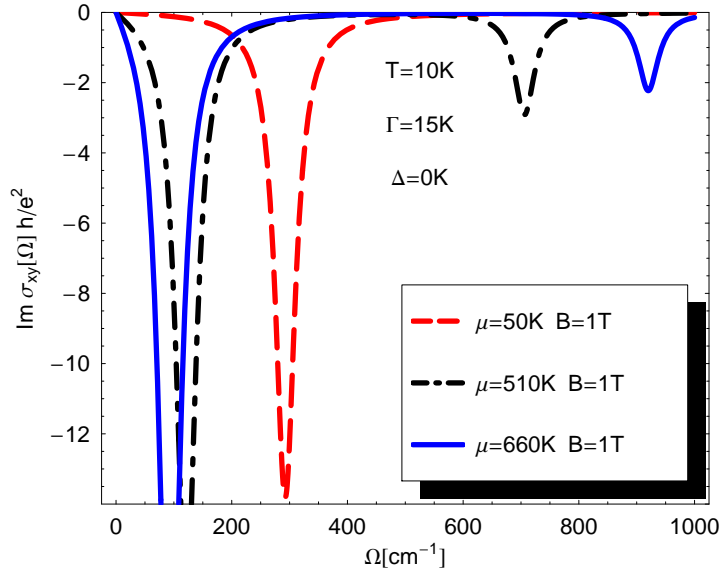}}
\caption{(Colour online) The imaginary part of the Hall
conductivity, $\mbox{Im} \, \sigma_{xy}(\Omega)$ in units of $e^2/h$
as a function of frequency $\Omega$ in $\mbox{cm}^{-1}$. The three
cases are for $\mu = 50 \mbox{K}$  (long dashed), $\mu = 510
\mbox{K}$ (dash-dotted) and $\mu = 660 \mbox{K}$ (solid). The other
parameters are $B=1 \mbox{T}$, $T=10 \mbox{K}$ and $\Gamma =15
\mbox{K}$. } \label{fig:8}
\end{figure}
In Fig.~\ref{fig:8} we show results for $\mbox{Im} \,
\sigma_{xy}(\Omega)$ in units of $e^2/h$ as a function of $\Omega$
in $\mbox{cm}^{-1}$. In all cases considered temperature $T=10
\mbox{K}$, impurity scattering rate $\Gamma=15 \mbox{K}$ and the
excitonic gap $\Delta=0$. The magnetic field $B=1 \mbox{T}$ and
three values of chemical potential are considered. The long dashed
(red) curve is for $\mu=50 \mbox{K}$ below the energy of the $n=1$
Landau level, dash-dotted (black) $\mu=510 \mbox{K}$ between $n=1$
and $n=2$, and solid (blue) $\mu=660 \mbox{K}$ between $n=2$ and
$n=3$ as in Fig.~\ref{fig:1}. We note a single peak in the long
dashed (red) curve at $M_1$ in contrast to two in the other two
curves. For the dash-dotted (black) curve the peaks are at $M_2-M_1$
and $M_1+M_2$, respectively, and for the solid (blue) curve they are
at $M_3-M_2$ and $M_2 +M_3$. These features can be easily understood
from Eq.~(\ref{sigma_xy-complex-corrected}) for $
\sigma_{xy}(\Omega)$ when we take its imaginary part in the limit
$\Delta=0$, $\Gamma \to 0$ which is for $\Omega>0$, $\mu>0$
\begin{eqnarray}
\label{Im_sigma_xy-Gamma=0}
\fl \mbox{Im} \,\sigma_{xy}(\Omega) = -\frac{e^2}{h} M_1^2
\frac{\pi}{2} \sum_{n=0}^{\infty}
[n_F(M_n)-n_F(M_{n+1})-n_F(-M_{n+1})+n_{F}(-M_n)]\\
\lo\times \left[\frac{\delta(\Omega - M_{n+1} + M_n)}{M_{n+1}-M_n} +
\frac{\delta(\Omega - M_{n+1} - M_n)}{M_{n+1}+M_n} \right].
\end{eqnarray}
Taking the limit of zero temperature, $T=0$ in
Eq.~(\ref{Im_sigma_xy-Gamma=0}) only the first two thermal factors
in the square bracket survive. For $\mu \in ]M_0,M_1[$ we get
\begin{equation}
\mbox{Im} \,\sigma_{xy}(\Omega) = -\frac{e^2}{h} M_1^2 \pi
\frac{\delta(\Omega-M_1)}{M_1},
\end{equation}
while for $\mu \in ]M_{N},M_{N+1}[$ with $N>0$  we get
\begin{equation}
\mbox{Im} \,\sigma_{xy}(\Omega) = -\frac{e^2}{h} M_1^2 \frac{\pi}{2}
\left[\frac{\delta(\Omega-M_{N+1} -M_{N})}{M_{N+1}+M_N} +
\frac{\delta(\Omega-M_{N+1}+M_{N})}{M_{N+1}-M_N}\right].
\end{equation}
Thus for $\mu \in ]M_0,M_1[$ the transverse Hall conductivity
exhibits a single peak at $\Omega = M_1$ with weight $2/M_1$, in
units of $(e^2/h) (\pi/2) M_1^2$  and for $\mu \in ]M_{N},M_{N+1}[$
with $N>0$ it has two peaks at $M_{N+1}+M_N$ and $M_{N+1}-M_N$ with
weight $1/(M_{N+1}+M_N)$ and $1/(M_{N+1}-M_N)$, respectively. Thus
the case $\mu \in ]M_0,M_1[$ is different from all others. This
distinguishes Dirac from a classical Landau level quantization, for
which, all cases would have two peaks. Also the area under the peaks
in $\mbox{Im}\, \sigma_{xy}(\Omega)$ is $\sim \sqrt{B}$ in analogy
to what we found for $\mbox{Re}\, \sigma_{xx}(\Omega)$.

Next we consider the effect of a finite excitonic gap on the
imaginary part of the Hall conductivity. In this case even for
$\Gamma \to 0$, $\Omega>0$, $\mu>0$ equation
(\ref{sigma_xy-complex-corrected}) is slightly more complicated than
(\ref{Im_sigma_xy-Gamma=0}) because of the additional factors $1 \pm
\Delta^2/M_nM_{n+1}$ which are only equal to $1$ when $\Delta=0$.
\begin{figure}[h]
\centering{
\includegraphics[width=10cm]{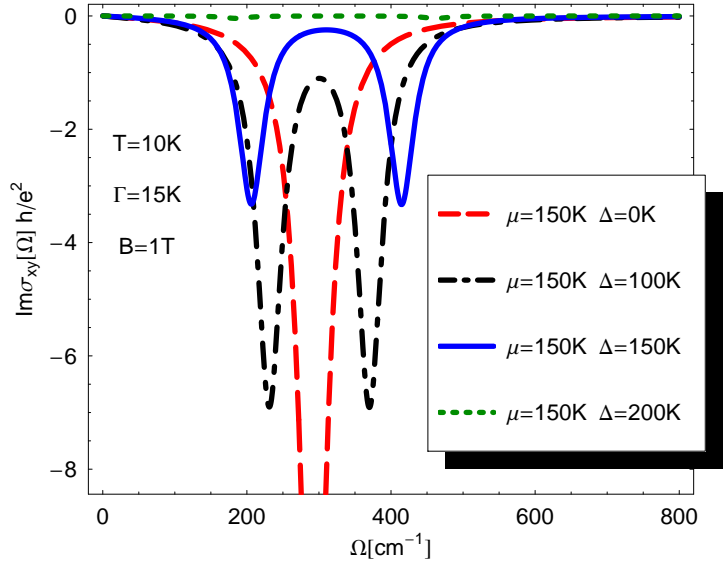}}
\caption{(Colour online) The imaginary part of the Hall
conductivity, $\mbox{Im} \,\sigma_{xy}(\Omega)$ in units of $e^2/h$
vs frequency $\Omega$ in $\mbox{cm}^{-1}$ for temperature $T=10
\mbox{K}$, $\Gamma =15 \mbox{K}$, $B=1 \mbox{T}$ and chemical
potential $\mu = 150 \mbox{K}$ for 4 values of the excitonic gap
$\Delta$. Long dashed $\Delta = 0 \mbox{K}$, dash-dotted $\Delta =
100 \mbox{K}$, solid $\Delta = 250 \mbox{K}$, short dashed $\Delta =
200 \mbox{K}$.} \label{fig:9}
\end{figure}
Nevertheless to understand physically the results given in
Fig.~\ref{fig:9} this extra complication is not needed and we can
use Eq.~(\ref{Im_sigma_xy-Gamma=0}) as a guide. What is most
important is to look at the thermal factors. We will consider only
the case when the chemical potential falls between $n=0$ and $n=1$
Landau level in energy. The relevant thermal factor is $n_F(M_0)
-n_F(M_1)$. All four curves in Fig.~\ref{fig:9} have $\mu=150
\mbox{K}$. The long dashed (red) curve is for $\Delta=0 \mbox{K}$
and is included for reference. It shows a single peak (in
$-\mbox{Im}\,\sigma_{xy}(\Omega)$) at $294 \mbox{cm}^{-1}$ as we
know from our previous discussion. For the dash-dotted (black) curve
$\Delta=100 \mbox{K}$ which falls below the value of chemical
potential (see Fig.~\ref{fig:2}~(b)). In this case the thermal
factors in Eq.~(\ref{Im_sigma_xy-Gamma=0}) at $T=0$ give $1$ and the
delta functions correspond to $\Omega = \sqrt{\Delta^2 + \Omega_1^2}
- \Delta$ and $\Omega =\sqrt{\Delta^2 + \Omega_1^2} + \Delta$. Thus,
the peak in the long dashed curve has split into two. Further such
peak has approximately, but not exactly the same optical spectral
weight as that under the single peak of long dashed line. This
arises because of the weighting factors $1\pm \Delta^2/M_n M_{n+1}$
of Eq.~(\ref{sigma_xy-complex-corrected}) not shown explicitly in
Eq.~(\ref{Im_sigma_xy-Gamma=0}). For $\Delta=\mu$ [solid (blue)
curve] the two peaks remain at $\Omega =\sqrt{\Delta^2 + \Omega_1^2}
\pm \Delta$ (note the small shift which corresponds to the slightly
different values of the gap between solid and short-long dashed
curve). However, the thermal factor $n_{F}(\Delta) -
n_F(\Delta+\sqrt{\Delta^2 + \Omega_1^2})$ now equals $1/2$ rather
than $1$ for the two previous cases, so that this feature on its own
reduces the optical spectral weight of these peaks by half. Finally
for the short dashed (green) curve the gap $\Delta$ is larger than
is the chemical potential and the thermal factor $n_F(M_0)
-n_F(M_1)$ is zero, so that no peak is seen.

\begin{figure}[h]
\centering{
\includegraphics[width=10cm]{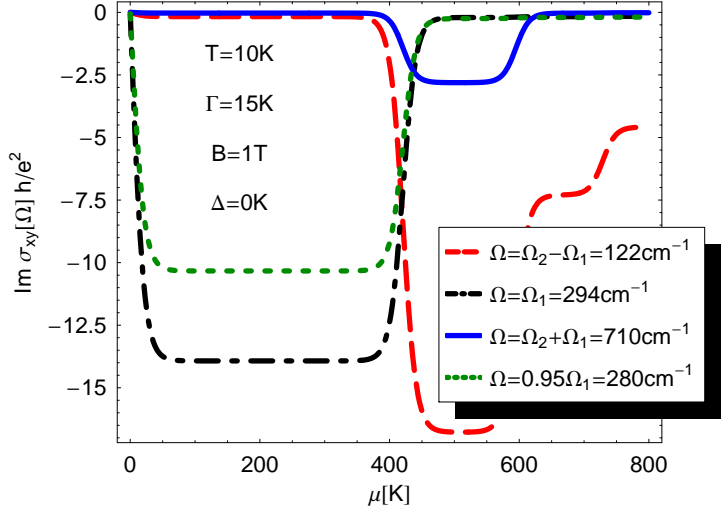}}
\caption{(Colour online) The imaginary part of the Hall
conductivity, $\mbox{Im} \,\sigma_{xy}(\Omega)$ in units of $e^2/h$
as a function chemical potential $\mu$ in $\mbox{K}$. Four
frequencies are considered, long dashed $\Omega = \Omega_2 -
\Omega_1 = 122 \mbox{cm}^{-1}$, dash-dotted $\Omega=\Omega_1 = 294
\mbox{cm}^{-1}$, solid $\Omega= \Omega_2 + \Omega_1 = 710
\mbox{cm}^{-1}$ and short dashed $\Omega = 0.95 \Omega_1=
280\mbox{cm}^{-1}$. The other parameters are $B=1 \mbox{T}$, $T=10
\mbox{K}$ and $\Gamma =15 \mbox{K}$. } \label{fig:10}
\end{figure}
In Fig.~\ref{fig:10} we show results for the change in $\mbox{Im} \,
\sigma_{xy}(\Omega)$ in units of $e^2/h$ as a function of $\mu$ in
Kelvin at fixed optical frequency. This figure is the analog of
Fig.~\ref{fig:5}. The same parameters $T=10 \mbox{K}$, $\Gamma=15
\mbox{K}$, $B=1 \mbox{T}$ and $\Delta=0 $ are chosen  as well as
$\Omega$, namely, long dashed (red) curve, $\Omega = \Omega_2 -
\Omega_1 = 122 \mbox{cm}^{-1}$. The absorptive Hall conductivity for
this curve is near zero until the first Landau level energy
$420\mbox{K}$ is crossed, where it drops below $-17e^2/h$ after
which it shows a plateau till the next level is crossed at $\mu=595
\mbox{K}$, where the second step is seen, etc. The dash-dotted
(black) curve is for $\Omega=\Omega_1 = 294 \mbox{cm}^{-1}$. In this
case the first plateau is at $\sim -14 e^2/h$ till the energy of the
$n=1$ Landau level is crossed in which case it drops to near zero
value. The short dashed (green) curve is for $\Omega =0.95 \Omega_1
= 280 \mbox{cm}^{-1}$ and follows the dash-dotted (black) curve
except the plateau is at $\sim -10 e^2/h$ as we expect. Finally the
solid (blue) curve is for $\Omega = \Omega_1 + \Omega_2 = 710
\mbox{cm}^{-1}$. It starts at zero till $\mu$ crosses $420 \mbox{K}$
at which point it shows a step down, remains nearly constant and
finally steps back to near zero value at $\mu=595 \mbox{K}$ as
expected from consideration of Fig.~\ref{fig:8}.

\begin{figure}[h]
\centering{
\includegraphics[width=10cm]{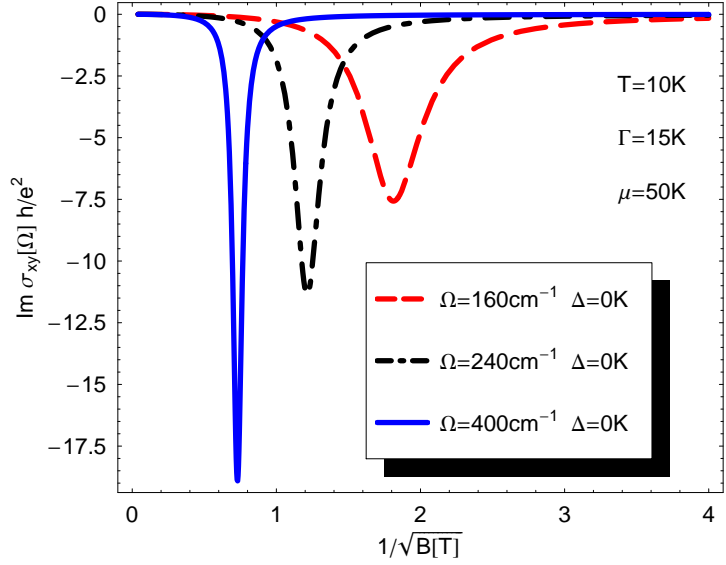}}
\caption{(Colour online) The imaginary part of the Hall
conductivity, $\mbox{Im} \, \sigma_{xy}(\Omega)$ in units of $e^2/h$
as a function of the inverse of the square root of the magnetic
field, $1/\sqrt{B}$ with $B$ in $\mbox{T}$. Three frequencies are
considered, long dashed $160 \mbox{cm}^{-1}$, dash-dotted $240
\mbox{cm}^{-1}$, solid $400 \mbox{cm}^{-1}$. The other parameters
are $T=10 \mbox{K}$, $\Gamma =15 \mbox{K}$ and $\mu=50 \mbox{K}$. }
\label{fig:11}
\end{figure}
The final Fig.~\ref{fig:11} shows results for the variation of
$\mbox{Im} \,\sigma_{xy}(\Omega)$ in units of $e^2/h$ vs the inverse
square root of the magnetic field, $1/\sqrt{B(T)}$. Here temperature
$T=10 \mbox{K}$, impurity scattering rate  $\Gamma= 15\, \mbox{K}$
and the chemical potential $\mu = 50 \mbox{K}$ with excitonic gap
$\Delta=0$. Three optical frequencies are chosen: long dashed (red)
curve $\Omega= 160 \mbox{cm}^{-1}$, dash-dotted (black) curve
$\Omega =240 \mbox{cm}^{-1}$ and solid (blue) curve $\Omega=400
\mbox{cm}^{-1}$. These curves are similar to those of
Fig.~\ref{fig:6} for the longitudinal case, but here  exhibit only a
single peak as is expected from the curves shown in
Fig.~\ref{fig:8}.

\section{Discussion}
\label{sec:concl}

In this paper we have extended the calculation of the ac
conductivity of graphene presented in Ref.~\cite{Gusynin2006PRB} in
several ways. One emphasis has been on presenting results for
different values of the chemical potential, $\mu$. In a field-effect
device, $\mu$ can be varied within the range of a few thousands of
Kelvin by changing the gate voltage
\cite{Geim2005Nature,Kim2005Nature,Novoselov2004Science,Bunch2005Nano}.
This implies that for fields of the order of $1 \mbox{T}$, $\mu$ can
be made to sweep over several Landau levels. Recently
\cite{Basov-organic} infrared spectroscopy has been successfully
applied to study FET devices based on poly(3-hexylthiophene). We
find a rich pattern of behaviour for the $\Omega$ dependence of the
real (absorptive) part of the diagonal conductivity, $\mbox{Re} \,
\sigma_{xx}(\Omega,T)$ as a function of $\Omega$ in a fixed external
magnetic field oriented perpendicular to the graphene sheet. Peaks
are seen in $\sigma_{xx}(\Omega,T)$ corresponding to the possible
transition between Landau levels from the hole to particle band
(interband) or within a given cone (intraband). For $\mu$ between
the $n=0$ and $n=1$ levels there are lines at $\Omega=M_1$,
$M_1+M_2$, $M_2 + M_3$ (interband), etc., where $M_i$ is the
position in energy of the $i$-th Landau level. There is no peak
below $M_1$. The relative optical spectral weight between the
various levels decay approximately as $1/\Omega$ with $\Omega$
evaluated at the center of each peak. When the chemical potential
falls between the energy of the $n=1$ and $n=2$ level the peak at
$\Omega= M_1$ fades into the background and a new line appears at
the lower frequency $M_2 - M_1$ (intraband). In addition, the
intensity of the line at $M_1 + M_2$ drops to half its value, while
all other lines remain the same. As the chemical potential crosses
higher and higher Landau level, say falls between $M_N$ and
$M_{N+1}$, the low energy peak has shifted to $M_{N+1} -M_N$
(intraband) after which all peaks previously seen have disappeared
into background except for the one at $M_{N+1}+M_{N}$ which  has
half its previous intensity. Again all peaks above this energy
remain unaltered. The peculiar behaviour of the peak at $\Omega=M_1$
which is either present with full intensity or completely absent is
the hallmark of Dirac as oppose to Schr\"odinger behaviour. This
peak can be classified either as inter or intra when the excitonic
gap $\Delta=0$, an ambiguity that is lifted for $\Delta\neq 0$. For
Landau levels based on the Schr\"odinger equation the interband
lines are evenly spaced while the intraband line is fixed in energy,
and all interband lines first half their intensity before
disappearing entirely as $\mu$ is increased.

We have found that the transition from one configuration of
absorption lines to another occurs for a small change in chemical
potential near a given Landau level energy with the scale for the
incremental change in $\mu$ set by temperature and/or level
broadening $\Gamma$. Away from these special values of $\mu$ the
curves do not change significantly.

In anticipation of experiments \cite{Basov-private} we also provide
scans of the the behaviour of $\mbox{Re} \, \sigma_{xx}(\Omega,T)$
at fixed $\Omega$ as a function of chemical potential or of magnetic
field. The pattern of behaviour found is traced back to that just
described for $\mbox{Re} \, \sigma_{xx}(\Omega,T)$ vs $\Omega$ at
different values of field and chemical potential.  Parallel results
for the absorptive part of the transverse Hall conductivity are also
presented.

The possibility that an excitonic gap may open in graphene under
high magnetic field has been considered by many authors
\cite{Khveshchenko2001PRL,Khveshchenko2001aPRL,Gorbar2002PRB,Gusynin2006catalysis}
and may even have been observed in recent experiments
\cite{Zhang2006PRL}. We have considered its effect on the absorption
peak seen in both diagonal and Hall conductivity. Our specific
predictions are that, a given peak can split into two, can disappear
entirely or can simply shift to higher energy without splitting
depending on the value of the chemical potential. In some
circumstances the optical spectral weight under the splitted peak
can differ from each other by a factor of order 3.

Many of the results shown in the figures were obtained on the basis
of general formulas for the conductivity, but it was found that
simplified versions which can be more easily used to interpret
experimental results are surprisingly accurate in the cases
considered. They involve sums over Landau level indices of
Lorentzian forms. Recently Li {\em et al.} \cite{Li2006} and
Sadowski {\em et al.} \cite{Sadowski2006} used related forms to
analyze their results.

Early magnetoreflectance data \cite{Toy1977PRB} in magnetic fields
in the range $1$ to $10 \, \mbox{T}$ carried out in graphite
assigned lines to the $H$-point and some of these lines were found
to follow a square root of $B$ law as expected for graphene. Recent
data of Ref.~\cite{Li2006} at higher fields up to $20\,\mbox{T}$,
however, found a conventional linear in $B$ behaviour. Very recent
infrared transmission data \cite{Sadowski2006} in ultrathin
epitaxial graphite \cite{Berger2004JPCB,Berger2006Science} in fields
up to $4 \, \mbox{T}$ do show $\sqrt{B}$ law and these authors took
this to be evidence that their carbon sheets are sufficiently
decoupled to behave like graphene. However, this interpretation also
requires that they assume that various sheets carry different
charges, i.e. have different values of chemical potential.
Nevertheless, the main lines seen in these data were assigned to the
first three Dirac interband transitions and the $(0,1)$ intraband
transition  on the basis of their position in energy. They were also
seen to vary in  optical spectral weight as the square root of the
magnetic field in good agreement with the findings in this work.
Many other detailed predictions made in this paper have yet to be
verified and should help in firmly establishing the special
characteristic of quasiparticles in graphene.

\ack We thank D.~Basov for sharing Ref.~\cite{Li2006} prior to
publication and E.J.~Nicol for discussion. The work of V.P.G. was
supported by the SCOPES-project IB7320-110848 of the Swiss NSF and
by Ukrainian State Foundation for Fundamental Research. J.P.C. and
S.G.Sh. were supported by the Natural Science and Engineering
Research Council of Canada (NSERC) and by the Canadian Institute for
Advanced Research (CIAR).

\appendix
\section{Calculation of $\sigma_{\pm}(\Omega)$ in magneto-optical Lorentzian model}
\label{sec:A}

We consider the complex conductivity
$\sigma_{\pm}(\Omega)=\sigma_{xx}(\Omega)\pm i\sigma_{xy}(\Omega)$.
Substituting the spectral function (\ref{spect-fun-magfield}) in
Eq.~(\ref{electric_cond}) after evaluating $\mbox{tr}$ and
integrating over momentum $\mathbf{k}$ [see Appendix~A of
Ref.~\cite{Gusynin2006PRB}] one obtains
\begin{equation}
\label{sigma_complex} \fl
\sigma_{\pm}(\Omega)=\frac{e^2N_fv_F^2|eB|}{4\pi c\,\Omega
i}\int_{-\infty}^{\infty} d\omega d\omega^\prime
\frac{n_F(\omega^\prime)-n_F(\omega)}{\omega-\omega^\prime-\Omega-i0}\left[\psi_1(\omega,\omega^\prime)
\mp {\rm sgn}(eB)\psi_2(\omega,\omega^\prime)\right],
\end{equation}
where the functions $\psi_1(\omega,\omega^\prime)$ and
$\psi_2(\omega,\omega^\prime)$ are:
\begin{equation}
\label{psi1-2topsi-n-m}
\psi_{1,2}(\omega,\omega^\prime)=\sum_{n,m=0}^\infty(-1)^{n+m+1}\left(\delta_{n,m-1}\pm\delta_{m,n-1}\right)
\psi_{n,m}(\omega,\omega^\prime)
\end{equation}
with
\begin{eqnarray}
\fl
\psi_{n,m}(\omega,\omega^\prime)=\left(1-\frac{\Delta^2}{M_nM_m}\right)\left(A_n(\omega)A_m(\omega^\prime)+
B_n(\omega)B_m(\omega^\prime)\right)\nonumber \\
\lo
+\left(1+\frac{\Delta^2}{M_nM_m}\right)\left(A_n(\omega)B_m(\omega^\prime)+
B_n(\omega)A_m(\omega^\prime)\right),
\end{eqnarray} and
\begin{equation}
A_n(\omega)=\frac{\Gamma_n}{\pi[(\omega-M_n)^2+\Gamma_n^2]},\qquad
B_n(\omega)=\frac{\Gamma_n} {\pi[(\omega+M_n)^2+\Gamma_n^2]}.
\end{equation}
One can easily check that \begin{equation}
\psi_{1}(\omega,\omega^\prime)=\psi_{1}(\omega^\prime,\omega),\quad
\psi_{2}(\omega,\omega^\prime)= -\psi_{2}(\omega^\prime,\omega).
\end{equation}
Using this symmetry one can rewrite Eq.~(\ref{sigma_complex}) in the
form
\begin{eqnarray} \label{sigma_pm}
\fl \sigma_{\pm}(\Omega)=-\frac{e^2N_fv_F^2|eB|}{4\pi c\,\Omega
i}\int_{-\infty}^{\infty} d\omega
\,n_F(\omega)\int_{-\infty}^{\infty}
d\omega^\prime\left\{\left[\frac{1}
{\omega-\omega^\prime+\Omega+i0}+\frac{1}{\omega-\omega^\prime-\Omega-i0}\right]
\psi_1(\omega,\omega^\prime)\right.\nonumber \\
\lo\pm\left.{\rm sgn}(eB)\left[\frac{1}
{\omega-\omega^\prime+\Omega+i0}-\frac{1}{\omega-\omega^\prime-\Omega-i0}\right]
\psi_{2}(\omega,\omega^\prime)\right\}.
\end{eqnarray}
Now using Eq.~(\ref{psi1-2topsi-n-m}) we obtain
\begin{eqnarray}
\fl \sigma_{\pm}(\Omega)=-\frac{e^2N_fv_F^2|eB|}{4\pi c\,\Omega
i}\sum_{n,m=0}^\infty(-1)^{n+m+1} \int_{-\infty}^{\infty}
d\omega\,n_F\left(\omega\right)\int_{-\infty}^{\infty}
d\omega^\prime \psi_{n,m}(\omega,\omega^\prime) \nonumber \\
\lo \times
\left\{\left[\frac{1}{\omega-\omega^\prime+\Omega+i0}+\frac{1}{\omega-\omega^\prime-\Omega-i0}\right]
\left(\delta_{n,m-1}+\delta_{m,n-1}\right)  \right. \nonumber \\
\left. \lo \pm {\rm sgn}(eB)\left[\frac{1}
{\omega-\omega^\prime+\Omega+i0}-\frac{1}{\omega-\omega^\prime-\Omega-i0}\right]
\left(\delta_{n,m-1}-\delta_{m,n-1}\right)\right\}.
\end{eqnarray}
For a typical integral over $\omega^\prime$ we have
\begin{eqnarray}
\fl  \int_{-\infty}^{\infty}
\frac{d\omega^\prime\psi_{n,m}(\omega,\omega^\prime)}
{\omega-\omega^\prime+\Omega+i0}=\left(1-\frac{\Delta^2}{M_nM_m}\right)\left(\frac{A_n(\omega)}
{\omega-M_m+\Omega+i\Gamma_m}+\frac{B_n(\omega)}
{\omega+M_m+\Omega+i\Gamma_m}\right)\nonumber \\
\lo
+\left(1+\frac{\Delta^2}{M_nM_m}\right)\left(\frac{B_n(\omega)}{\omega-M_m+\Omega+i\Gamma_m}+
\frac{A_n(\omega)}{\omega+M_m+\Omega+i\Gamma_m}\right).
\end{eqnarray}
If the thermal factor $n_F$ is absent, further integration over
$\omega$ would give an exact result:
\begin{eqnarray}
\fl \int_{-\infty}^{\infty} d\omega d\omega^\prime
\frac{\psi_{n,m}(\omega,\omega^\prime)}{\omega-\omega^\prime+\Omega+i0}= \\
\fl \left(1-\frac{\Delta^2}{M_nM_m}\right)\left(\frac{1}
{M_n-M_m+\Omega+i(\Gamma_n+\Gamma_m)}+\frac{1}{-M_n+M_m+\Omega+i(\Gamma_n+\Gamma_m)}\right)\nonumber \\
\fl + \left(1+\frac{\Delta^2}{M_nM_m}\right)\left(\frac{1}
{-M_n-M_m+\Omega+i(\Gamma_n+\Gamma_m)}+\frac{1}{M_n+M_m+\Omega+i(\Gamma_n+\Gamma_m)}\right).
\nonumber
\end{eqnarray}
In case we integrate with a smooth function $n_F\left(\omega\right)$
we can approximately write
\begin{eqnarray}
\label{approx1}
\fl \int_{-\infty}^{\infty}  d\omega\,n_F\left(\omega\right)
\hspace{-2mm}\int_{-\infty}^{\infty}
\frac{d\omega^\prime\psi_{n,m}(\omega,\omega^\prime)}
{\omega-\omega^\prime+\Omega+i0}\simeq \\
\fl
\left(1-\frac{\Delta^2}{M_nM_m}\right)\left(\frac{n_F\left(M_n\right)}
{M_n-M_m+\Omega+i(\Gamma_n+\Gamma_m)}+\frac{n_F\left(-M_n\right)}
{-M_n+M_m+\Omega+i(\Gamma_n+\Gamma_m)}\right)\nonumber\\
\fl+\left(1+\frac{\Delta^2}{M_nM_m}\right)\left(\frac{n_F\left(-M_n\right)}{-M_n-M_m+\Omega+i(\Gamma_n+\Gamma_m)}+
\frac{n_F\left(M_n\right)}{M_n+M_m+\Omega+i(\Gamma_n+\Gamma_m)}\right)
\nonumber.
\end{eqnarray}
Similarly, taking the complex conjugate and changing
$\Omega\to-\Omega$, we have
\begin{eqnarray}
\label{approx2}
\fl \int_{-\infty}^{\infty}  d\omega\,n_F\left(\omega\right)
\int_{-\infty}^{\infty}
\frac{d\omega^\prime\psi_{n,m}(\omega,\omega^\prime)}
{\omega-\omega^\prime-\Omega-i0} \simeq  \\
\fl
\left(1-\frac{\Delta^2}{M_nM_m}\right)\left(\frac{n_F\left(M_n\right)}
{M_n-M_m-\Omega-i(\Gamma_n+\Gamma_m)}+\frac{n_F\left(-M_n\right)}
{-M_n+M_m-\Omega-i(\Gamma_n+\Gamma_m)}\right)\nonumber \\
\fl
+\left(1+\frac{\Delta^2}{M_nM_m}\right)\left(\frac{n_F\left(-M_n\right)}
{-M_n-M_m-\Omega-i(\Gamma_n+\Gamma_m)}+
\frac{n_F\left(M_n\right)}{M_n+M_m-\Omega-i(\Gamma_n+\Gamma_m)}\right).
\nonumber
\end{eqnarray}
Hence we arrive at
\begin{eqnarray}
\label{Lorentz-full}
\fl \sigma_{\pm}(\Omega)=-\frac{e^2N_fv_F^2|eB|}{4\pi c\,\Omega i}\\
\fl \times \sum_{n=0}^{\infty}
\left\{\left(1-\frac{\Delta^2}{M_nM_{n+1}}\right)\left[[n_F\left(M_n\right)-n_F\left(M_{n+1}\right)]
+[n_F\left(-M_{n+1}\right)-n_F\left(-M_{n}\right)]\right]\right.\nonumber \\
\lo \times\left.\left[\frac{1}{M_n-M_{n+1}+
\Omega+i(\Gamma_n+\Gamma_{n+1})}+\frac{1}{M_n-M_{n+1}-\Omega-i(\Gamma_n+\Gamma_{n+1})}\right]
\right.\nonumber \\
\lo-\left.\left(1+\frac{\Delta^2}{M_nM_{n+1}}\right)\left[[n_F\left(-M_{n+1}\right)-n_F\left(M_{n}\right)]
+[n_F\left(-M_{n}\right)-n_F\left(M_{n+1}\right)]\right]\right.\nonumber\\
\lo \times\left.\left[\frac{1}{M_n+M_{n+1}-
\Omega-i(\Gamma_n+\Gamma_{n+1})}+\frac{1}{M_n+M_{n+1}+\Omega+i(\Gamma_n+\Gamma_{n+1})}\right]
\right.\nonumber \\
\fl \pm{\rm
sgn}(eB)\left.\left[\left(1-\frac{\Delta^2}{M_nM_{n+1}}\right)\left[[n_F\left(M_n\right)-n_F\left(M_{n+1}\right)]
-[n_F\left(-M_{n+1}\right)-n_F\left(-M_{n}\right)]\right]\right.\right. \nonumber \\
\lo \times\left.\left.\left[\frac{1}{M_n-M_{n+1}+
\Omega+i(\Gamma_n+\Gamma_{n+1})}-\frac{1}{M_n-M_{n+1}-\Omega-i(\Gamma_n+\Gamma_{n+1})}\right]
\right.\right.\nonumber \\
\lo+\left.\left.\left(1+\frac{\Delta^2}{M_nM_{n+1}}\right)\left[[n_F\left(-M_{n+1}\right)-n_F\left(M_{n}\right)]
-[n_F\left(-M_{n}\right)-n_F\left(M_{n+1}\right)]\right]\right.\right.\nonumber \\
\lo\times\left.\left.\left[\frac{1}{M_n+M_{n+1}-
\Omega-i(\Gamma_n+\Gamma_{n+1})}-\frac{1}{M_n+M_{n+1}+\Omega+i(\Gamma_n+\Gamma_{n+1})}\right]\right]\right\},
\nonumber
\end{eqnarray}
where only the sum over $n$ remained.

We verified that $\mbox{Re} \, \sigma_{xx}(\Omega)$ computed for
$\Gamma_{n}(\omega) = \mbox{const}$ from the more approximate
Eq.~(\ref{Lorentz-full}) agrees quantitatively with the results
obtained from a full Eq.~(\ref{optical-diagonal}). This agreement is
best for the resonance peaks and only small deviations are seen for
$\Omega \sim 0$. Nevertheless, Eq.~(\ref{Lorentz-full}) has a few
drawbacks due to approximations made in Eqs.~(\ref{approx1}),
(\ref{approx2}).

In particular, the Drude form cannot be recovered in $B \to 0$
limit, while it can be obtained
\cite{Gusynin2006PRB,Gusynin2006micro} from an exact representation
(\ref{optical-diagonal}). Moreover, the imaginary parts of the
diagonal conductivity, $\mbox{Im} \sigma_{xx}(\Omega)$ and the Hall
conductivity, $\mbox{Im}\sigma_{xy}(\Omega)$ are divergent in the
limit $\Omega \to 0$ and do not satisfy Kramers-Kronig relations
with the corresponding real parts found from
Eq.~(\ref{Lorentz-full}). To correct these problems we move the term
$1/\Omega$ under the sum replacing it by its value at the pole of
the corresponding denominator in Eq.~(\ref{Lorentz-full}) and arrive
at Eqs.~(\ref{sigma_xx-complex-corrected}) and
(\ref{sigma_xy-complex-corrected}).

\end{document}